\documentclass[prd,10pt,nofootinbib]{revtex4-1}
\usepackage[latin1]{inputenc}

\usepackage{amsmath}
\usepackage{amsfonts}
\usepackage{mathrsfs}
\usepackage{amssymb}
\usepackage{graphicx}
\usepackage{geometry}
\usepackage{hyperref}
\usepackage{xcolor}
\usepackage{float}

\newgeometry{vmargin={25mm}, hmargin={20mm}}

\newcommand{\s}[1]{\textrm{#1}}
\newcommand{\st}[1]{\textrm{\tiny#1}}
\newcommand{\E}[0]{\mathcal{E}}
\newcommand{\sv}{\langle \sigma_\s{ann} v \rangle}

\begin{document}
	\title{Equilibrium axisymmetric halo model for the Milky Way and its implications for direct and indirect DM searches}
	\author{Mihael Peta\v c}
	\email{petac@lupm.in2p3.fr}
	\affiliation{Laboratoire Univers et Particules de Montpellier (LUPM) \& CNRS \& Universit\'e de Montpellier (UMR-5299), Place Eug\`ene Bataillon, F-34095 Montpellier Cedex 05, France}
	\
	\date{\today}
	
	\begin{abstract}
	We for the first time provide self-consistent axisymmetric phase-space distribution models for the Milky Way's dark matter (DM) halo which are carefully matched against the latest kinematic measurements through Bayesian analysis. By using broad priors on the individual galactic components, we derive conservative estimates for the astrophysical factors entering the interpretation of direct and indirect DM searches. While the resulting DM density profiles are in good agreement with previous studies, implying $\rho_\odot \approx 10^{-2} \, M_\odot / \s{pc}^3$, the presence of baryonic disc leads to significant differences in the local DM velocity distribution in comparison with the standard halo model. For direct detection, this implies roughly 30\% stronger cross-section limits at DM masses near detectors maximum sensitivity and up to an order of magnitude weaker limits at the lower end of the mass range. Furthermore, by performing Monte-Carlo simulations for the upcoming DARWIN and DarkSide-20k experiments, we demonstrate that upon successful detection of heavy DM with coupling just below the current limits, the carefully constructed axisymmetric models can eliminate bias and reduce uncertainties by more then 50\% in the reconstructed DM coupling and mass, but also help in a more reliable determination of the scattering operator. Furthermore, the velocity anisotropies induced by the baryonic disc can lead to significantly larger annual modulation amplitude and sizable differences in the directional distribution of the expected DM-induced events. For indirect searches, we provide the differential $J$-factors and compute several moments of the relative velocity distribution that are needed for predicting the rate of velocity-dependent annihilations. However, we find that accurate predictions are still hindered by large uncertainties regarding the DM distribution near the galactic center.
	\end{abstract}

	\maketitle
	
	\tableofcontents
	
	\section{Introduction}
	\label{sec:introduction}
	
	Throughout its history, the Dark Matter (DM) hypothesis has been successful in explaining a series of independent observations which span from the galactic up to cosmological scales~\cite{bretone_history_2018}. Also within our own galaxy, various kinematic tracers of the gravitational potential have shown the need for a massive DM halo~\cite{oort_note_1960,bahcall_giants_1984,kuijken_galactic_1991,fich_rotation_1989,merrifield_rotation_1992,dehnen_mass_1998,salucci_dark_2010,sofue_rotation_2013,catena_novel_2010,iocco_evidence_2015,read_local_2014}. The latter has been long recognized as a promising target for the discovery of DM particles, either through direct detection in lab experiments~\cite{goodman_detectability_1985} or by indirect searches aimed at detecting emissions related to DM annihilation or decay~\cite{zeldovich_astrophysical_1980,silk_gamma-ray_1987}. On the other hand, the existing experiments have not yet provided an undisputed detection of DM signals. Instead, robust upper limits on the strength of possible interactions between DM and Standard Model (SM) particles have been obtained~\cite{jianglai_current_2017,conrad_indirect_2017,schumann_direct_2019}. Its coupling to baryons has been extensively probed by direct detection experiments which use large loads of various target materials to search for nuclear recoils induced by their scattering with the local DM. At the same time, independent limits regarding DM's annihilation into SM states have been established from indirect searches where particularly strong bounds come from the absence of excess radiation, consistent with DM explanation, in the central part of the Milky Way~\cite{abazajian_strong_2020}. However, all of these constraints crucially depend on the modeling of DM distribution within our galaxy. For instance, the nuclear recoil rate, which is the key observable in direct detection experiments, is proportional to the local DM density multiplied by a convolution of the relevant cross-section with DM's velocity distribution.
	An accurate modeling of local DM is also needed for correctly predicting various signatures of DM induced events which are crucial for rejecting backgrounds.
	On the other hand, in indirect searches the annihilation flux is proportional to the DM density squared integrated along the line of sight, which makes the bounds particularly sensitive to the DM distribution near the galactic center. In the simplest case of s-wave annihilations the DM velocity distribution plays no role, however, it becomes important in numerous well-motivated DM models where non-trivial velocity dependences of the annihilation cross-section arise -- see, e.g.,~\cite{hagelin_perhaps_1984,kang_singlet_2008,pospelov_secluded_2008,tulin_beyond_2013,das_selection_2017,dent_thermal_2010,iengo_sommerfeld_2009,slatyer_sommerfeld_2010}.
	
	In the past crude models of the galactic halo were often used, typically relying on uncorrelated estimates of DM density and velocity dispersion while assuming Maxwellian velocity distribution.
	While such approximations were well justified in the early studies, we have today several reasons to go beyond such simple modeling. Most notably, the quality of astronomical observations has significantly improved within the last decade, which allows one to successfully constrain more realistic models of the galactic halo that are based on dynamical equilibrium. An unprecedented amount of information regarding the structure and dynamics of the Milky Way was recently obtained by precise astrometric measurements of the Gaia mission~\cite{gaia_mission_2016}. However, accurate complementary observations, such as stellar spectroscopy~\cite{majewski_apache_2017} and variability sky surveys~\cite{wegg_moa-ii_2016,mroz_rotation_2019}, also provided important new insights. At the same time, the increasing sensitivity of direct and indirect searches calls for improvements in the theoretical predictions of the expected DM signals, which intricately depend on the galactic DM distribution. As already mentioned, accurate halo modeling is particularly important in the interpretation of direct detection experiments and will become even more so in the future since several characteristic features of DM-induced events, such as their energy spectrum, annual modulation and directional distribution, are highly sensitive to the modeling of DM velocity distribution.
	Furthermore, many diverse explanations regarding the nature of DM have been suggested throughout the literature, and in order to encompass such a wide theoretical landscape, as well as correctly reconstruct the properties of DM particles upon their detection, a precise model for the structure of the galactic DM distribution is indispensable.
	
	Guided by these considerations, we obtain new results regarding the equilibrium phase-space distribution of DM particles within the Milky Way. Such attempts were made previously in the context of Eddington's inversion method which, however, relies on spherical symmetry. Instead, we deploy a generalized inversion method that is applicable to axisymmetric systems and, therefore, provides a significantly better description of our galaxy, since the baryonic disc is knwon to dominate the dynamics within the inner $\sim 10 \; \mathrm{kpc}$.
	Following the Bayesian approach, we use Monte-Carlo Markov Chain (MCMC) sampling to simultaneously fit the DM distribution as well as the baryonic gravitational potential to a set of the latest kinematic constraints. This, in turn, allows us to self-consistently compute equilibrium two-integral phase-space distribution function (PSDF) for the galactic DM through a computationally efficient method previously explored in~\cite{hunter_two-integral_1993,qian_axisymmetric_1995,petac_two-integral_2019}. It is important to note that our approach provides an accurate description only for DM particles that have reached dynamical equilibrium. While the latter should be the case for the majority of galactic DM, a certain fraction of non-equilibrium structures is expected to be present, such as tidal debris from past mergers~\cite{meyong_sausage_2018,meyong_milky_2018,helmi_merger_2018,necib_inferred_2019} or ongoing accretion of smaller objects~\cite{freese_effects_2004,ohare_dark_2018,necib_evidence_2019}. These can often be conveniently expressed as corrections to the underlying equilibrium distribution -- see, e.g.,~\cite{evans_shm$++$:_2018,ibarra_impact_2019,necib_evidence_2019,buch_implications_2020}. By assuming that all of the galactic DM is in dynamical equilibrium (while the re-scaling to a different fraction of the equilibrium component is trivial), we compute key astrophysical quantities that enter the interpretation of direct and indirect DM searches and consistently propagate the associated errors stemming from the kinematic observations. Using these results, we highlight the key advantages of our approach in interpreting direct as well as indirect searches for DM. We also provide tabulated values for all the relevant astrophysical factors, making them particularly convenient for the use in future analyses of experimental results.
	
	In section~\ref{sec:DM_modeling}, we begin with a review of conventional approaches for modeling the galactic DM distribution. This is followed by a summary of the axisymmetric inversion method and our assumptions regarding the DM density and halo rotation. In section~\ref{sec:observations} we briefly summarize the existing constraints on the structure of our galaxy and present the compilation of observational data that will be used in our subsequent analysis. The latter is described in section~\ref{sec:mass_depomposition}, where we also present the results obtained from the sampling of our benchmark galactic models. In section~\ref{sec:direct_detection}, we turn our attention to the predictions for DM-induced signals in direct detection experiments, where we pay particular attention to discrepancies with respect to the standard halo model. The implications of our DM halo models for indirect searches are presented in section~\ref{sec:indirect_detection}. We summarize our results and state our conclusions in section~\ref{sec:conclusions}.
	
	\section{modeling of the galactic DM}
	\label{sec:DM_modeling}
	
	Various observations provide compelling evidence that DM halos are composed of self-gravitating, non-relativistic and (at least to a good approximation) collisionless particles. Such a system can be conveniently described by a phase-space distribution function, $f$, which measures the number of particles per phase-space volume:
	\begin{align}
	\mathrm{d}N = f(\vec{p},\vec{q}) \; \mathrm{d}^3p \; \mathrm{d}^3q \; ,
	\end{align}
	where $\vec{p}$ and $\vec{q}$ is the relevant pair of canonical coordinates. For galactic DM searches, phase-space distribution functions are particularly useful since they contain all information regarding the system, i.e. the spatial and velocity distribution of DM particles. The evolution of $f$ is governed by the Boltzmann equation for which equilibrium solutions can be obtained under certain simplifying assumptions. However, the full power of such rigorous approach has only been appreciated recently, while significantly simpler models had been used in the past. In the interpretation of direct DM searches, as well as indirect searches focusing on velocity-dependent annihilations, it was often assumed that $f$ is separable in independent spatial and velocity distributions and can be written as:
	\begin{align}
	\label{eqn:shm_psdf}
	f(r, v) = \rho(r) P(v) \; ,
	\end{align}
	where $\rho(r)$ is the DM density profile and $P(v)$ the DM velocity distribution. In the context of direct detection, most of the theoretical estimates of the expected DM-induced signals relied on the Standard Halo Model (SHM), which for the local DM assumes a truncated Maxwellian velocity distribution:
	\begin{align}
	P(v) & = \mathcal{N} \exp \left(-\frac{v^2}{2 \sigma^2} \right) \Theta \left(v_\s{esc} - v\right) \; , \\
	& \textrm{where} \;\;\; \mathcal{N}^{-1} = \left(2 \pi \sigma^2 \right)^{3/2} \left( \textrm{erf} \left(\frac{v_\s{esc}}{\sqrt{2 \sigma^2}} \right) - \sqrt{\frac{2}{\pi}} \frac{v_\s{esc}}{\sigma} \exp \left( - \frac{v^2_\s{esc}}{2 \sigma^2} \right) \right) \; . \nonumber
	\end{align}
	The above ansatz is parameterized by two quantities, namely the DM velocity dispersion $\sigma$ and the local escape velocity $v_\s{esc}$, which are both challenging to constrain from observations. Their estimates can only be obtained by adopting some concrete mass model of our galaxy and matching it against various kinematic tracers of the galactic gravitational potential. It also tacitly assumes isotropic velocity dispersion, however, this is most probably not true for the local DM, given that the flattened baryonic disc dominates the dynamics within the Milky Way even beyond the solar galactocentric distance. The possibility of anisotropic Maxwellian velocity distribution was recently considered in~\cite{bozorgnia_dark_2019}. However, in this case, additional assumptions regarding the velocity dispersion tensor are needed since its components are otherwise impossible to constrain through the available observations. A further drawback of the SHM lies in the fact that it treats the local DM density and the parameters entering $P(v)$ as independent, while by adopting a concrete galactic model and matching it against a set of kinematic tracers intricate correlations among these quantities typically arise~\cite{lavalle_making_2015,green_astrophysical_2017}. Finally, it should be noted that the SHM represents an equilibrium configuration only if the DM density profile corresponds to the one of the isothermal sphere, i.e. $\rho(r) \propto r^{-2}$, while the dynamics within the Milky Way, as well as cosmological simulations, suggest an appreciable different DM density distribution.
	
	To overcome the shortcomings of the SHM, several works advocated the use of Eddington's inversion formula~\cite{vergados_new_2003,catena_local_2012,lacroix_anatomy_2018,lacroix_predicting_2020}. The latter provides a simple way of obtaining unique stationary phase-space distribution functions for collisionless particles with a given radial density profile, $\rho(r)$, embedded in a spherical gravitational potential $\Psi(r)$~\footnote{Throughout this work $\Psi(\vec{r})$ denotes the relative gravitational potential which is defined as $\Psi(\vec{r}) \equiv -\Phi(\vec{r}) + \Phi_0$, where $\Phi(\vec{r})$ is the standard gravitational potential and $\Phi_0$ a constant such that $\Psi(\vec{r})$ vanishes at the boundary of the system. Since the Milky Way can be fairly well approximated as an isolated object, we set $\Phi_0 = 0$. In the case of a truncated system, such as satellite galaxies, equation~\eqref{eqn:eddington}, as well as~\eqref{eqn:psdf} and~\eqref{eqn:psdf_odd}, contain additional boundary term -- for more detailed discussion see, e.g.,~\cite{lacroix_anatomy_2018,petac_velocity-dependent_2018}.}:
	\begin{align}\label{eqn:eddington}
	f_\s{Edd}(\E) = \frac{1}{\sqrt{8} \pi^2} \cdot \frac{\mathrm{d}}{\mathrm{d} \E} \int_0^\E \frac{\mathrm{d} \Psi}{\sqrt{\E - \Psi}} \cdot \frac{\mathrm{d} \rho}{\mathrm{d} \Psi} \; .
	\end{align}
	In the above expression, $\E \equiv \Psi(r) - v^2/2$ is the relative energy that fully parameterizes $f$ in the case of ergodic (i.e. spherically symmetric and isotropic) systems. While Eddington's inversion allows one to obtain the DM phase-space distribution that is consistent with the presence of additional galactic components that enter by contributing to the total gravitational potential, it is limited to spherically symmetric systems. Furthermore, in its original formulation it can be used to reconstruct only isotropic distribution functions. Some progress has been recently made in the direction of anisotropic configurations~\cite{bozorgnia_anisotropic_2013,fornasa_self-consistent_2014,vergados_dark_2016,lacroix_anatomy_2018}, but only under further assumptions regarding the anisotropy profile of the studied particles. In this work, we instead use a generalization of the Eddington's approach, which can be applied to axisymmetric systems. As such, it is capable of providing a much more accurate description of disc galaxies and capturing the effects of a flattened baryonic component on the velocity distribution of DM particles. It also allows one to address additional features of DM halos that are often found in numerical simulations of structure formation, such as oblateness and rotation, which can not be self-consistently incorporated in spherically symmetric models. While the rotational properties of the Milky Way's halo remain uncertain, recent advances in the quality of astronomic observations allow us to constrain the distribution of baryons and DM with unprecedented precision. Therefore, we find it timely to revise the galactic mass decomposition and provide unique constraints on the axisymmetric phase-space distribution of DM in the light of newly available data. In the following, we will first briefly present the numerically efficient method of obtaining the relevant two-integral distribution function which will be followed by a short discussion of our assumptions regarding the parametric form of the DM density profile and rotational properties of the galactic DM halo.
	
	\subsection{Two-integral distribution function}
	\label{sec:HQ_method}
	
	Throughout this work, we will use the standard cylindrical coordinates, $(R, \phi, z)$, and assume that the galactic halo has no other continuous symmetries apart from the rotation around the central axis. The latter implies that the PSDF can be written as a function of two integrals of motion, namely $f = f(\E, L_z)$, where $\E$ is the aforementioned relative energy and $L_z$ is the angular momentum around the axis of symmetry, i.e. $L_z = R v_\phi$. In this case, a generalization of the Eddington's inversion formula can be obtained, allowing one to compute $f(\E, L_z)$ for an arbitrary axisymmetric density-potential pair. In particular, we will adopt a numerically friendly approach developed by Hunter \& Qian~\cite{hunter_two-integral_1993,qian_axisymmetric_1995}, that relies on theoretical foundations previously laid out by Lynden-Bell~\cite{lynden-bell_stellar_1962}. Until recently, the method was applied only to stellar systems, while it was for the first time systematically studied in the context of DM in~\cite{petac_two-integral_2019}. We refer the reader to these previous publications for the proof of the method and detailed discussions regarding its implementation, while our code for computing the PSDF is publicly available online.~\footnote{ \url{https://github.com/mpetac/AIM}} In the remainder of this section, we will provide a short summary of the method.
	
	Under the assumptions stated above, the PSDF can be decomposed in two parts, $f_+$ that is even in $L_z$ and $f_-$ that is odd:
	\begin{align}
	f(\E, L_z) = f_+(\E, |L_z|) + f_-(\E, L_z) \; ,
	\end{align}
	The even part contains information regarding the density distribution, while the odd part describes the rotational properties of the considered system. Hunter \& Qian~\cite{hunter_two-integral_1993,qian_axisymmetric_1995} showed that the $L_z$-even part of PSDF can be computed by providing an analytic continuation of the density-potential pair in the complex plane and evaluating the following contour integral:
	\begin{align} \label{eqn:psdf}
	f_+(\E, |L_z|) = \frac{1}{4 \pi^2 i \sqrt{2}} \oint_{C(\E)} \frac{\mathrm{d} \xi}{\sqrt{\xi - \E}} \left. \frac{\mathrm{d}^2 \rho(R^2, \Psi)}{\mathrm{d} \Psi^2} \right|_{\substack{\Psi = \xi \;\;\;\;\;\;\;\;\;\; \\ R^2=\frac{L_z^2}{2(\xi - \E)}}} \; .
	\end{align}    
	In the above expression $C(\E)$ refers to a path which tightly wraps around the real axis between the value of the potential at spatial infinity and a value corresponding to a circular orbit with relative energy $\E$, while $\rho$ is considered as a function of the radial coordinate and the total gravitational potential which is in principle always possible for monotonic $\Psi(R ,z)$. However, in great majority of practical situations one cannot express the density profile as an explicit function of the total gravitational potential and one is forced to perform the derivative implicitly using cylindrical coordinates:
	\begin{align} \label{eqn:derivative_expansion}
	\frac{\mathrm{d}^2 \rho(R^2, \Psi)}{\mathrm{d} \Psi^2} & =  \frac{\mathrm{d}^2 \rho(R^2, z^2)}{\mathrm{d} (z^2)^2} \left(\frac{\mathrm{d} \Psi(R^2, z^2)}{\mathrm{d} z^2} \right)^{-2} \nonumber \\ & \;\;\;\;\;\;\;\;\;\;
	- \frac{\mathrm{d} \rho(R^2, z^2)}{\mathrm{d} z^2} \frac{\mathrm{d}^2 \Psi(R^2, z^2)}{\mathrm{d} (z^2)^2}\left(\frac{\mathrm{d} \Psi(R^2, z^2)}{\mathrm{d} z^2} \right)^{-3} \; ,
	\end{align}
	and evaluate it at $R^2=\frac{L^2}{2(\xi - \E)}$ and $z^2$ such that $\Psi(R^2, z^2) = \xi$. Values of $z^2$ fulfilling the latter equality typically need to be found via numerical minimization routines. The $L_z$-odd part of PSDF can be computed analogously, using the following expression:
	\begin{align} \label{eqn:psdf_odd}
	f_-(\E, L_z) = \frac{\textrm{sign}(L_z)}{8 \pi^2 i} \oint_{C(\E)} \frac{\mathrm{d} \xi}{\xi - \E} \left. \frac{\mathrm{d}^2 \left( \rho \bar{v}_\phi \right)}{\mathrm{d} \Psi^2} \right|_{\substack{\Psi = \xi \;\;\;\;\;\;\;\;\;\; \\ R^2=\frac{L_z^2}{2(\xi - \E)}}} \; .
	\end{align}
	It is important to note that in order to evaluate $f_-$ one needs to specify also the rotation profile, $\bar{v}_\phi(R^2,z^2)$, which is, unfortunately, in the case of the galactic DM halo very poorly constrained. In analogy to the Eddington's inversion, the above described method can be in principle used to compute the PSDF for any choice of axisymmetric $\rho(R^2, z^2)$, $\Psi(R^2, z^2)$ and $\bar{v}_\phi(R^2, z^2)$. However, there is no guarantee that the resulting distribution function will be positive definite (i.e. physical). This needs to be verified explicitly after performing the inversion.
	
	\subsection{Parametrization of DM density and halo rotation}
	\label{sec:DM_assumptions}
	
	In our analysis we will consider two benchmark density profiles, namely the simulation-motivated NFW profile~\cite{navarro_universal_1997} with a central cusp:
	\begin{align} \label{eqn:nfw}
		\rho\st{NFW}(r) = \frac{\rho_s}{r / r_s \cdot (1 + r / r_s)^2} \; ,
	\end{align}
	and Burkert profile~\cite{burkert_structure_1995} with a central core:
	\begin{align} \label{eqn:burkert}
		\rho_\s{Bur}(r) = \frac{\rho_s}{(1 + r / r_s) \cdot (1 + r^2 / r_s^2)} \; ,
	\end{align}
	which are both parameterized by the DM scale density $\rho_s$ and scale length $r_s$. The above spherical density profiles can be generalized to the case of ellipsoidal halos by substituting $r \rightarrow m = \sqrt{R^2 + z^2 / q^2}$, where $q$ is the flattening parameter (i.e. $q < 1$ corresponds to oblate, $q > 1$ to prolate and $q = 1$ to spherical halo). However, several recent studies suggest that the Milky Way's halo is consistent with $q = 1$~\cite{bovy_shape_2016,williams_run:_2017,malhan_constraining_2019,wegg_gravitational_2019,nitschai_first_2019} and, therefore, we will in this work avoid addressing the possibility of ellipsoidal DM halo. As a side benefit, in the case of spherical NFW or Burkert halo it is possible to obtain analytical expressions for its gravitational potential, which greatly reduces the computational cost of obtaining the PSDF function through the axisymmetric inversion method. Besides oblateness, one could also consider more general parametric forms for the DM density distribution. While we do not expect this would have significant impact on the local DM distribution, which is of the prime interest for direct detection experiments, assumptions regarding the central DM density slope can have important implications for the interpretation of indirect searches. However, accurately determining the DM distribution around the galactic center is extremely challenging due to the lack of reliable observations as well as subdominant contribution of DM to the dynamics within the region. Therefore, we will in this work restrain from analyzing the very central part of our galaxy and rather focus on the global DM distribution which is much better constrained by the available kinematic data.
	
	For non-rotating DM halos the $L_z$-odd part of the PSDF vanishes, but for rotating halos one needs to additionally specify its rotation profile and compute $f_-(\E, L_z)$. Unfortunately, there exist no observational constraints regarding the rotational properties of the Milky Way's halo and one can study them only through comparisons with numerical simulations. Most of DM halos in hydrodynamic simulations exhibit some degree of net rotation, but this strongly depends on the particular merger history of an individual object. In general, different approaches of modeling $f_-(\E, L_z)$ are possible, e.g., making it proportional to $f_+(\E, |L_z|)$, or computing it through equation~\eqref{eqn:psdf_odd} by assuming a particular functional form for $\bar{v}_\phi(R, z)$. In this work we decided for the second option, since it assures more realistic rotational properties, by adopting a simple parametric rotation profile:
	\begin{align}
	\label{eqn:rotation_profile}
	\bar{v}_\phi(R) = \frac{\omega R}{1 + R^2 / r_a^2} \; .
	\end{align}
	In the above expression the characteristic radius, $r_a$, was chosen to correspond to the scale radius of the DM halo, i.e. $r_a = r_s$, while $\omega$ was tuned to reproduce the typical spin parameter of comparable galaxies found in hydrodynamical simulations. We define the spin parameter as~\cite{bullock_universal_2001}:
	\begin{align}
		\label{eqn:spin_parameter}
		\lambda(r) = \frac{J(r)}{\sqrt{2} r M(r) V_c(r)} \; ,
	\end{align}
	where $J(r)$ and $M(r)$ are the total angular momentum and mass of the DM halo within a sphere of radius $r$, while $V_c(r)$ is the velocity of a circular orbit at the given radius. As our benchmark value we adopt $\lambda(0.25 \; r_{200}) = 0.04$, where $r_{200}$ is the radius at which the average enclosed DM density equals to 200 times the mean cosmological DM density. The chosen value of $\lambda$ corresponds to the median value of spin parameter found by~\cite{bryan_impact_2013} and is consistent with the results of other recent studies, see, e.g.,~\cite{genel_galactic_2015,zavala_link_2016,zjupa_angular_2017} and references therein.
	
	\section{Observational constraints}
	\label{sec:observations}
	
	It has been long recognized that the Milky Way is composed of a significant amount of non-luminous matter~\cite{bretone_history_2018}. The existence of these early hints in favor of the DM hypothesis is perhaps not surprising since a broad range of kinematic tracers of the galactic gravitational potential is observationally accessible within our galaxy. Independent estimates on the amount and distribution of DM within the Milky Way have been obtained from the galactic rotation curve~\cite{salucci_dark_2010,weber_determination_2010,catena_novel_2010,mcmillan_mass_2011,catena_local_2012,bovy_local_2012,nesti_dark_2013,sofue_rotation_2013,iocco_evidence_2015,sofue_rotation_2017,mcmillan_mass_2017,eilers_circular_2019}, dynamics of halo stars~\cite{xue_milky_2008,deason_broken_2012,kafle_shoulders_2014,huang_milky_2016}, globular clusters~\cite{eadie_cumulative_2019,watkins_evidence_2019,posti_mass_2019} or satellite galaxies~\cite{peebles_dynamics_2017,callingham_mass_2019}, the distribution of hyper-velocity stars~\cite{piffl_rave_2014,fragione_constraining_2017,rossi_joint_2017,williams_run:_2017,monari_escape_2018,deason_local_2019} and orbits of tidal streams~\cite{gibbons_skinny_2014,bovy_shape_2016,malhan_constraining_2019}. In this work, we will follow the approach of global galactic mass decomposition based on the Milky Way's rotation curve and vertical motion of stars. We first divide the total mass of our galaxy in distinct components, for which parametric functions based on their characteristic morphology will be assumed, and subsequently match the model against kinematic data. For the latter, we rely on the latest determinations of rotation curve and vertical force above the galactic disc, which both benefited from the recent data release of the Gaia mission~\cite{collaboration_gaia_2018}. We chose the particular combination of constraints since they provide complementary information regarding the variation of the gravitational potential, namely in the directions along and perpendicular to the baryonic disc, which is crucial for properly constraining axisymmetric models. Additionally, the rotation curve, as well as the vertical force measurements, can be extracted from the observations under very modest assumptions and, therefore, provide a robust way of bracketing the distribution of DM within the Milky Way.
	In the following, we first turn our attention to the assumptions regarding the parametric functions used for approximating the baryonic components and the associated priors since they can have a significant impact on results of the galactic mass decomposition approach, as was recently demonstrated in~\cite{salas_estimation_2019}. This will be followed by an overview of recent studies of the galactic rotation curve and vertical force field, which will be used later on in our analysis.

	\subsection{Distribution of baryons}
	\label{sec:baryon_modeling}
	
	Despite significant improvements in observational data, the distribution of baryons within the Milky Way is still subjected to sizable uncertainties. This is primarily due to the fact that we are viewing our galaxy from within, which makes it harder to resolve the extent and precise shape of the stellar disc as well as the structure of our galaxy near to its center. In the following, we will present the baryonic model that will be used throughout our analysis, consisting of a spherical bulge and an axisymmetric disc. Even though this setup is rather simplistic, we believe it is sufficient to capture the shape of the baryonic gravitational potential and provide robust constraints on the considered DM density profiles. For a more complete review regarding the structure of our galaxy see, e.g.,~\cite{bland-hawthorn_galaxy_2016,wegg_moa-ii_2016,pouliasis_milky_2017,bienayme_new_2018,bland-hawthorn_galah_2019} and references therein.
	
	The central region of our galaxy is composed of a barred stellar bulge which exhibits intricate triaxial structure~\cite{clarke_milky_2019} that falls beyond the scope of axisymmetric models. In recent studies it has been often approximated by a spheroidal power-law distribution, flattened along the central axis, with exponential truncation in the outskirts~\cite{bissantz_spiral_2002,bland-hawthorn_galaxy_2016}. However, since the truncation radius has been estimated to lie at $r_\s{trunc} \sim 2$ kpc, which roughly coincides with the innermost determinations of the galactic rotation curve, the exact details of its morphology are not important for the analysis at hand. Therefore, we find it sufficient to approximate its gravitational potential using the spherical Hernquist ansatz~\cite{hernquist_analytical_1990}:
	\begin{align} \label{eqn:potential_H}
	\Psi_\st{H}(R^2, z^2) = \frac{G M_\s{bulge}}{\sqrt{R^2 + z^2} + a_\s{bulge}} \; ,
	\end{align}
	with appropriately small scale radius, i.e. $a_\s{bulge} \lesssim 1$ kpc, leaving its total mass, $M_\s{bulge}$, as the only free parameter. Regarding the latter there are significant uncertainties which mainly arise due to its overlap with the galactic disc. The strongest constraints on the stellar mass within the bulge region come from microlensing observations which are sensitive to the optical depth in the central part of our galaxy. Most of the recent studies point towards $M_\s{bulge} \sim 1.4 - 1.8 \cdot 10^{10} M_\odot$~\cite{portail_made--measure_2015,wegg_moa-ii_2016} which, however, include also the disc stars and therefore smaller masses, e.g. $M_\s{bulge} \sim 10^{10} M_\odot$~\cite{mcmillan_mass_2017,pouliasis_milky_2017}, have been often used in combination with appropriate disc models. In our attempt to marginalize over the baryonic uncertainties we will only use an upper prior on the bulge mass, i.e. $M_\s{bulge} < 1.8 \cdot 10^{10} M_\odot$, while keeping the scale length fixed to $a_\s{bulge} = 0.5$ kpc.
	
	Regarding the galactic disc, there are several caveats which make it difficult to build a constrained mass model without relying on rather strong assumptions. First of all, the baryonic disc is known to deviate from axial symmetry due to the presence of complex features, such as spiral arms or warps caused by external gravitational perturbations. However, such detailed modeling of the galactic disc falls well beyond the scope of this work and, therefore, we will rely on the standard assumption of axisymmetric disc. Throughout the literature, the presence of distinct thin and thick stellar discs was often assumed, both following a double-exponential density profile:
	\begin{align} \label{eqn:exp_disc}
	\rho_\s{disc} = \frac{\Sigma_0}{2 z_\s{disc}} \exp \left(-\frac{R}{R_\s{disc}} - \frac{|z|}{z_\s{disc}}\right) \; ,
	\end{align}
	where $R_d$ and $z_d$ are the scale length and scale height, while $\Sigma_0$ is the corresponding surface density.  The thick disc is often approximated by scale height $z_\s{disc}^\s{thick} \approx 0.9$ kpc and relatively small scale length, $R_\s{disc}^\s{thick} \approx 2.5$ kpc, while the thin disc was often modeled as having $z_\s{disc}^\s{thin} \approx 0.3$ kpc and $R_\s{disc}^\s{thin} \approx 3.5$ kpc, see, e.g.,~\cite{bland-hawthorn_galaxy_2016} and references therein. The ratio between their local surface densities was estimated as $f_\Sigma \sim 0.12$ with rather large uncertainties due to their overlap, with the thin disc being the dominant one. These definitions are based on the spatial distribution and kinematic properties of the galactic stars, assuming the double-exponential morphology. However, several recent studies have shown that it is more appropriate to talk about discs composed of $\alpha$-rich and $\alpha$-poor stars~\cite{bovy_spatial_2012,hayden_chemical_2015,bovy_stellar_2016,bland-hawthorn_galah_2019}, since such classification is not subjected to ambiguities despite the significant overlap of the two components. Unfortunately, these new definitions of the stellar discs are not fully consistent with the aforementioned model, and some caution is needed. The observed $\alpha$-rich disc is appreciably thicker, $z_\s{disc}^{\alpha\s{-rich}} \sim 1$ kpc, and can be traced up to the solar radius, while the $\alpha$-poor disc has a larger but poorly constrained radial extent and varying height, with $z_\s{disc}^{\alpha\s{-poor}} \sim 0.3$ kpc in the central part and slowly increasing towards the outskirts where it ``flares" and reaches thickness comparable or even greater than the $\alpha$-rich disc. It has also been argued that both of the discs have roughly the same mass~\cite{pouliasis_milky_2017}, however, due to large uncertainties regarding their scale lengths also their total masses can not be very accurately determined. Furthermore, it has been pointed out by several authors that even more distinct populations can be identified by looking at other element ratios within the stars. Due to the lack of a coherent picture and well-constrained disc parameters we chose to model the galactic disc as a single Miyamoto-Nagai component (in which we also include the subdominant contribution of gas which was traditionally modeled as an additional double-exponential disc) whose gravitational potential takes the following analytical form~\cite{miyamoto_three-dimensional_1975}:
	\begin{align} \label{eqn:potential_MN}
	\Psi_\st{MN}(R^2, z^2) = \frac{G M_\s{disc}}{\sqrt{R^2 + (a_\s{disc} + \sqrt{z^2 + b_\s{disc}^2})^2}} \; ,
	\end{align}
	The validity of such modeling is supported by the fact that combining several exponential discs with different scale lengths and scale heights leads to an overall gravitational potential which can be fairly well approximated by the MN expression.~\footnote{We explicitly checked that the combination of the standard double-exponential thin and thick discs yields a gravitational potential which can be well approximated by the Miyamoto-Nagai ansatz. More precisely, one can find values of $M_\s{disc}$, $a_\s{disc}$ and $b_\s{disc}$ that provide a match with relative accuracy better than 15\% for any value of $R$ and $z$. An even better match can be found by restricting to $R \gtrsim 2$ kpc, which is the relevant range for present analysis -- in this case, the deviations can be reduced to less than 10\%.} Since in this work we are primarily interested in bracketing the uncertainties related to baryonic distribution and not reconstructing individual disc components, such modeling performs sufficiently well. At the same time, it greatly simplifies the analysis by introducing fewer free parameters and allows expressing the corresponding gravitational potential in an analytical form, while in the case of double-exponential density~\eqref{eqn:exp_disc} the latter needs to be computed through numerical quadrature. Regarding its mass, we will use only an upper bound $M_\s{disc} < 10^{11} M_\odot$, see, e.g.,~\cite{bland-hawthorn_galaxy_2016}, which is in fact much larger than the total mass of traditional baryonic disc components. Similarly, we will use highly agnostic priors for the disc scale length and scale height, namely $a_\s{disc} < 6$ kpc and $b_\s{disc} < 1$ kpc, which are again very generous bounds with respect to the values found in the literature. However, to break the degeneracy between the disc mass and scale length we adopt recently updated values of the local surface density of baryons $\Sigma_\odot = 47.1 \pm 3.4 \; M_\odot / \s{pc}^2$~\cite{mckee_stars_2015}, which is based on stellar counts and gas mapping within the solar neighborhood and does not rely on any assumption regarding the morphology, size and number of baryonic discs.
	
	\subsection{Milky Way's rotation curve}
	\label{sec:rotation_curve}
	
\begin{figure}[h]
		\centering
		\includegraphics[width=5in]{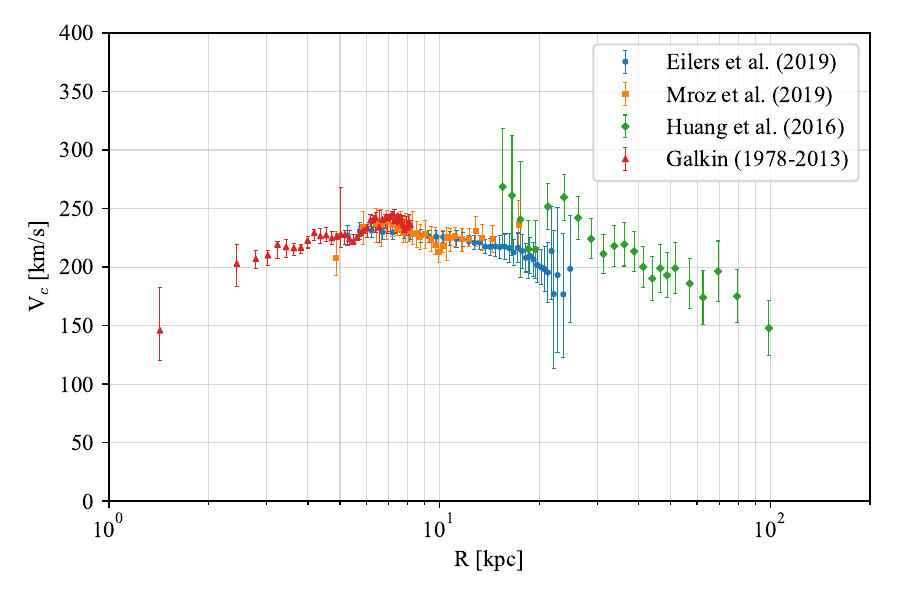}
		\caption{Compilation of Milky Way's circular velocity determinations, that we use in our analysis, as a function of the galactocentric distance.}
		\label{fig:vc_data}
	\end{figure}

	For galactic mass decomposition, particularly valuable information comes from the measurements of stellar and gas circular velocities, $V_c(R)$. The latter is related to the radial variation of the total gravitational potential in the galactic plane:
	\begin{align} \label{eqn:circular_velocity}
	V_c^2(R) = - R \cdot \left. \frac{\partial \Psi_\s{tot}(R,z)}{\partial R} \right|_{z=0} \; .
	\end{align}
	Recently significant progress in the observations allowed for more accurate determination of stellar kinematics within the Milky Way. The main improvement comes from the precise astrometric data provided by the Gaia mission, which mapped positions and proper motions for an overwhelming number of stars. Relying on these measurements, accurate determinations of the galactic rotation curve were obtained from a large sample of red giant stars as well as variable Cepheid stars, for which accurate complementary distance measurements are available. While these studies presently provide the best determination of $V_c(R)$ for galactocentric distances in the range from 8 to 25 kpc, additional information at larger galactocentric distances is required to constrain the DM halo's scale radius successfully. For this reason somewhat older determinations of circular velocities from halo giant stars, which were mapped up to a distance of 100 kpc by SDSS-III/SEGUE survey, also play an important role. While it is possible to obtain estimates on the enclosed galactic mass, $M(r)$, at even larger galactocentric distances (where $V^2_c(R) \approx G (M(r) / r)|_{r=R}$) based on the proper motions of the Milky Way's satellite galaxies~\cite{peebles_dynamics_2017,collaboration_gaia_GC_dsph_2018,callingham_mass_2019,fritz_mass_2020}, we chose to restrain from using them in our analysis because of two reasons. Firstly, such constraints strongly depend on the assumptions used for modeling the trajectories of the satellite galaxies and, secondly, they are subjected to significantly larger uncertainties than the local $V_c(R)$ determinations, which makes them irrelevant for constraining the DM content within the inner part of the Milky Way, even when assuming rigid DM density profiles, such as the ones used in this work. On the other hand, at small radii valuable measurements of the galactic rotation curve can be gained from the terminal velocities of gas, which provide the most accurate measurements of $V_c(R)$ at radii within the solar galactocentric distance. According to the above discussion, we chose to include in our analysis the following studies of the Milky Way's rotation curve:
	\begin{enumerate}
		\item Eilers et al. (2019)~\cite{eilers_circular_2019}: observations of 23000 red-giants stars in the range of 5 to 25 kpc.
		
		\item Mroz et al. (2019)~\cite{mroz_rotation_2019}: observations of 773 Classical Cepheids in the range of 5 to 20 kpc.
		
		\item Huang et al. (2016)~\cite{huang_milky_2016}: observations of halo K-giant stars in the range from 16 to 100 kpc.
		
		\item Galkin (1978-2013)~\cite{pato_$textbackslashtextttgalkin$:_2017}: compilation of rotation curve determinations based on terminal velocities of gas and masers in the range from 1.4 to 8 kpc.
	\end{enumerate}
	While Eilers et al. and Huang et al. provide binned data, Mroz et al. estimate $V_c$ for each star separately and, therefore, their results have to be binned to be on the same footing as the other two studies. We also chose to bin the Galkin data, since the complete compilation contains several distinct surveys with uncompetitively large error bars. To perform the binning, we split the relevant datasets in $\sqrt{N}$ bins, where $N$ is the total number of data points, chosen such that each bin contains roughly the same number of elements, and computed the median, 16th and 84th percentile of each bin, as the corresponding central value and its errors. At this point we also note that Eilers et al. and Mroz et al. used in their analysis the recent determination of the solar galactocentric distance $R_\odot = 8.122 \pm 0.031$ kpc~\cite{collaboration_detection_2018}.~\footnote{There authors later published a work with an updated value of the galactocentric distance of the Sun, $R_\odot = 8.178 \pm 0.013 \; \s{kpc}$~\cite{gravity_geometric_2019} Since this change in $R_\odot$ is very small with respect to the observational errors on the kinematic data, it should not have a significant effect on the derived rotation curve and, subsequently, our results. In the remainder of this work we will, therefore, assume $R_\odot = 8.122$ kpc.}, while Huang et al. used somewhat older value of $R_\odot = 8.34 \pm 0.16$ kpc. Strictly speaking, the rotation curve derived by the latter is not consistent with the other two. However, this should not significantly affect our analysis since the error bars for the rotation curve provided by Huang et al. are much larger and dominate over the difference that stems from the adopted value of $R_\odot$. On the other hand, rotation curves obtained from Galkin were rescaled to $R_\odot = 8.122$ kpc, local circular velocity of $V_c(R_\odot) = 231$ km/s~\cite{eilers_circular_2019,mroz_rotation_2019} and peculiar motion of the Sun $(U_\odot, V_\odot, W_\odot) = (11.1, 12.24, 7.25)$ km/s~\cite{schoenrich_local_2010} to be brought in agreement with the values used by Eilers et al. and Moroz et al. The final set of $V_c(R)$ data points is shown in figure~\ref{fig:vc_data}.
	
	\begin{figure}[h]
		\centering
		\includegraphics[width=0.8\textwidth]{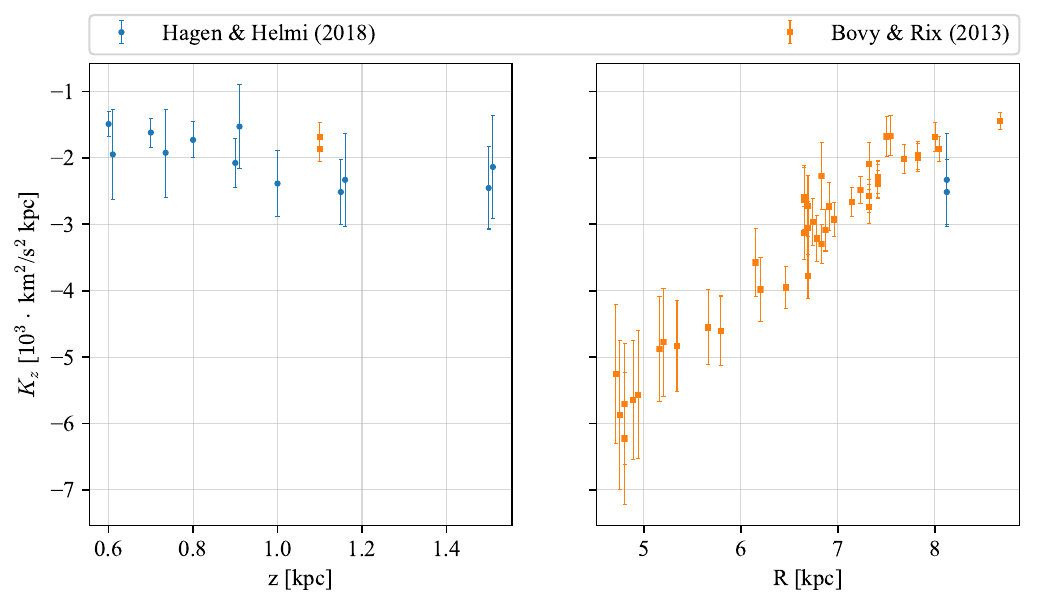}
		\caption{Vertical force estimates that we use in our analysis. The left-hand side plot shows $K_z$ as a function of height above the galactic midplane at $R \sim R_\odot$, while the right-hand side plot shows $K_z$ as a function of radial distance at $z \sim 1.1$ kpc.}
		\label{fig:kz}
	\end{figure}
	
	\subsection{Vertical motion of stars}
	\label{sec:vertical_motion}
	
	The vertical motion of stars has been long known to provide a powerful tool for constraining the DM density in the solar neighborhood -- see, e.g., ~\cite{bahcall_k_1984,kuijken_galactic_1991,creze_distribution_1998,silverwood_non-parametric_2016}. Most analyses rely on the axisymmetric Jeans modeling~\cite{binney_galactic_2008} which relates the observed stellar distribution and kinematics to the underlying gravitational potential. By singling out the vertical direction (i.e. the one perpendicular to the galactic disc), the following relation can be obtained:
	\begin{align} \label{eqn:vertical_motion}
	\frac{1}{\nu} \frac{\partial}{\partial z} (\nu \sigma^2_z) + \frac{1}{R \nu} \frac{\partial}{\partial R} \left( R \nu \sigma^2_{Rz} \right) = \frac{\partial \Psi_\s{tot}(R, z)}{\partial z} \; ,
	\end{align}
	where $\nu(R, z)$ is the number density of the tracer stars, $\sigma_{z}(R, z)$ is their velocity dispersion along the $\hat{z}$-axis and $\sigma_{Rz}(R, z)$ is the off-diagonal element of the velocity dispersion tensor corresponding to the meridional plane, i.e. $\sigma_{Rz} = \langle (v_R - \bar{v}_R) (v_z - \bar{v}_z) \rangle$. As can be seen by comparing equations~\eqref{eqn:circular_velocity} and~\eqref{eqn:vertical_motion}, the observations of circular velocity and vertical motion of stars probe the galactic gravitational potential in orthogonal directions, as the former is proportional to its derivative along the $\hat{R}$ and the latter along $\hat{z}$ coordinate, and, therefore, provide valuable complementary information. In particular, the vertical motion is very useful in breaking the degeneracy between spherically symmetric and flattened mass components that typically occurs when relying exclusively on rotation curve data. Furthermore, it allows us to obtain tighter constraints on the flattening of the gravitational potential which is crucial for accurate reconstruction of the DM's PSDF through the axisymmetric inversion method.
	There have been several recent studies utilizing the accurate Gaia astrometry to constrain the local mass distribution along the vertical direction, either focusing on the thin stellar populations \cite{buch_using_2019,widmark_measuring_2019} that span up to few hundred pc above the disc, as well as thick populations \cite{hagen_vertical_2018,sivertsson_local_2018} that probe the galactic potential up to $z \approx 1.5$ kpc.
	In our work we chose to use the results of Hagen \& Helmi \cite{hagen_vertical_2018} derived from the motion of thick disc stars, as their findings are largely consistent with other works, but they also directly provide the inferred vertical force $K_z(R,z) = \partial \Psi_\s{tot}(R, z) / \partial z$. On the other hand, we do not include the studies focused on thin disc only since they are somewhat controversial, finding evidence for an additional highly flattened component, which is speculated to be either an underestimated gaseous disc~\cite{widmark_measuring_2019} or even a dark disc~\cite{buch_using_2019}, but could also be due to the departure from equilibrium which is tacitly assumed by the Jeans analysis; for the accumulating evidence in favor of the latter see, e.g.,~\cite{antoja_dynamically_2018,khoperskov_echo_2019,bennett_vertical_2019,bland-hawthorn_galah_2019}. An alternative approach of studying the vertical force is to use the full 6D stellar phase-space information. This has been done by Bovy \& Rix \cite{bovy_direct_2013}, who derived $K_z(R,z=1.1$ kpc$)$ at various galactocentric radii in the range from $4-9$ kpc, however, using pre-Gaia observations. We chose to include their determinations of $K_z(R,z) $ since they are consistent with newer studies and allow us to constrain better the scale length of the galactic disc. The final compilation of $K_z(R,z)$ measurements used in the analysis is shown in figure \ref{fig:kz}.
	
	\section{Galactic mass decomposition}
	\label{sec:mass_depomposition}
	
	Computing the PSDF of DM using the axisymmetric inversion method, presented in section~\ref{sec:HQ_method}, requires the knowledge of the DM density distribution as well as the total gravitational potential of the galaxy. Various observations of gas and stars in the Milky Way provide us with information regarding the amount of baryons and their spatial distribution within the galaxy, while their kinematics can be used to constrain the over-all gravitational potential. By combining this information one can reconstruct the missing dynamical mass which must be sourced by the DM halo. Formally, this can be expressed through the Poisson equation, which relates the sum of densities corresponding to individual galactic components, $\rho_i(\vec{r})$, to the Laplacian of the total gravitational potential:
	\begin{align} \label{eqn:poisson}
	\nabla^2 \Psi_\s{tot}(\vec{r}) = - 4 \pi G \sum_i \rho_i(\vec{r}) \; .
	\end{align}
	According to the discussion in section~\ref{sec:baryon_modeling}, we will parameterize the gravitational potential of baryonic components using the Miyamoto-Nagai ansatz~\eqref{eqn:potential_MN} for the disc and Hernquist ansatz~\eqref{eqn:potential_H} for the bulge. For the DM halo, we will consider two qualitatively different parametric density functions, namely the NFW and Burkert profile, given by equations~\eqref{eqn:nfw} and~\eqref{eqn:burkert} respectively. The superposition of these galactic components will be then constrained by circular velocity and vertical force determinations, presented in sections~\ref{sec:rotation_curve} and~\ref{sec:vertical_motion}, together with the complementary measurement of the local baryon surface density.
	
	\subsection{Bayesian sampling of the galactic mass models}
	
	In order to constrain the DM distribution within the Milky Way, we will make use of the Bayesian approach based on Monte-Carlo Markov Chain (MCMC) exploration of the parameter space, given the set of constraints discussed above. Our benchmark galactic models have in total 6 free parameters:
	\begin{align}
	\vec{\theta} = \{M_\s{bulge}, M_\s{disc}, a_\s{disc}, b_\s{disc}, \rho_s, r_s \} \; ,
	\end{align}
	where the first four are related to the baryonic distribution while the last two describe the DM density profile. For all the baryonic parameters we adopt broad non-informative (i.e. flat) priors, which can be safely established according to the existing literature, as described in section~\ref{sec:baryon_modeling}:
	\begin{eqnarray}
	0 \; \leq & M_\s{bulge} & \leq 1.8 \cdot 10^{10} M_\odot \; , \nonumber \\
	0 \; \leq & M_\s{disc} & \leq 10^{11} M_\odot \; , \nonumber \\
	0 \; \leq & a_\s{disc} & \leq 6 \; \s{kpc} \; , \nonumber \\
	0 \; \leq & b_\s{disc} & \leq 1 \; \s{kpc} \nonumber
	\end{eqnarray}
	Regarding the DM parameters, we adopt the most generous range of priors, i.e. demanding that $\rho_s$ and $r_s$ are non-negative and impose that the scale radius is within the range of available data:
	\begin{align}
	& 0 < \rho_s \nonumber \\
	& 0 < r_s < 100 \; \s{kpc}
	\end{align}
	Since the scale density can generally vary over several orders of magnitude we chose to sample it using a logarithmic variable $\tilde{\rho}_s \equiv \log_{10} \left( \rho_s / M_\odot \s{pc}^{-3} \right)$. The 6-dimensional parameter space was then explored according to the generalized Gaussian likelihood that is capable of accommodating asymmetric errors of the data, $\mathcal{D}$. By assuming that the errors follow a split normal distribution one can use the following likelihood function (for details see~\cite{barlow_asymmetric_2003,villani_multivariate_2006}):
	\begin{align}
	\label{eqn:loglikehood_mw}
	\mathcal{L}(\mathcal{D}|\vec{\theta})  \equiv & \prod_{k = 1}^{N_{V_c}} 
	\frac{2}{\sqrt{2 \pi } \, (\sigma^+_{V_{c,k}} + \sigma^-_{V_{c,k}})}
	\exp \left[ -\frac{1}{2} \left( \frac{ V_{c, k} - V_c(R_k; \vec{\theta}) }{\sigma^\pm_{V_{c,k}}} \right)^2 \right]  \nonumber \\
	& \cdot \prod_{l = 1}^{N_{K_z}} 
	\frac{2}{\sqrt{2 \pi } \, (\sigma^+_{{K_{z,l}}} + \sigma^-_{{K_{z,l}}})}
	\exp \left[ -\frac{1}{2} \left( \frac{ K_{z, l} - K_z(R_l, z_l; \vec{\theta}) }{\sigma^\pm_{K_{z,l}}} \right)^2 \right] \; \nonumber \\
	& \cdot \frac{1}{\sqrt{2 \pi } \, \sigma_{\Sigma_\odot}} \exp \left[ -\frac{1}{2} \left( \frac{ \Sigma_\odot - \Sigma(R_\odot; \vec{\theta}) }{\sigma_{\Sigma_\odot}} \right)^2 \right]
	\end{align}
	In the first two lines of the above expression $V_{c,i}$ ($K_{z,i}$) are the binned rotation curve (vertical force) measurements, with total of $N_{V_c}$ ($N_{K_z}$) points, and $\sigma^\pm_{V_{c,i}}$ ($\sigma^\pm_{K_{z,i}}$) the corresponding lower/upper error estimates, while $V_c(R_i; \vec{\theta})$ ($K_z(R_i, z_i; \vec{\theta})$) are the predictions of our model at radial distance of the bin $R_i$ (and vertical height $z_i$) for given parameter vector $\vec{\theta}$. Since the observational errors on the circular velocity and vertical force are asymmetric, the upper error $\sigma^+_{V_{c,i}/K_{z,i}}$ should be used if the measured central value is below the theoretical prediction and $\sigma^-_{V_{c,i}/K_{z,i}}$ if it is above. In the last line of the above expression $\Sigma_\odot$ is the observationally inferred local baryon surface density and $\sigma_{\Sigma_\odot}$ its corresponding standard deviation, while $\Sigma(R_\odot; \vec{\theta})$ is the prediction of our model. For the exploration of parameter space we relied on Python implementation of Goodman \& Weare's affine invariant MCMC Ensemble sampler, delivered in the \textit{emcee} package~\cite{foreman-mackey_emcee_2013}. The sampling was done using 200 walkers, where each of them evolved for 20000 steps. The first half of each chain was discarded as part of the burn-in period.
	
	\subsection{Results}
	\label{sec:mass_decomposition_results}
	
	\begin{figure}[p]
		\centering
		\includegraphics[width=0.49\textwidth]{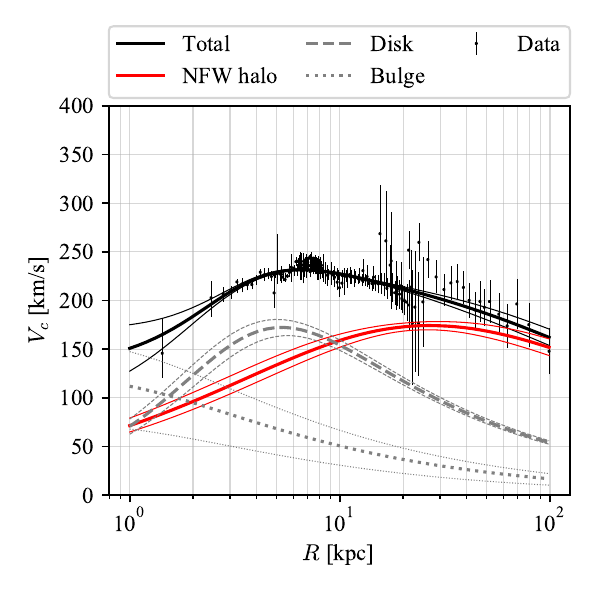}
		\includegraphics[width=0.49\textwidth]{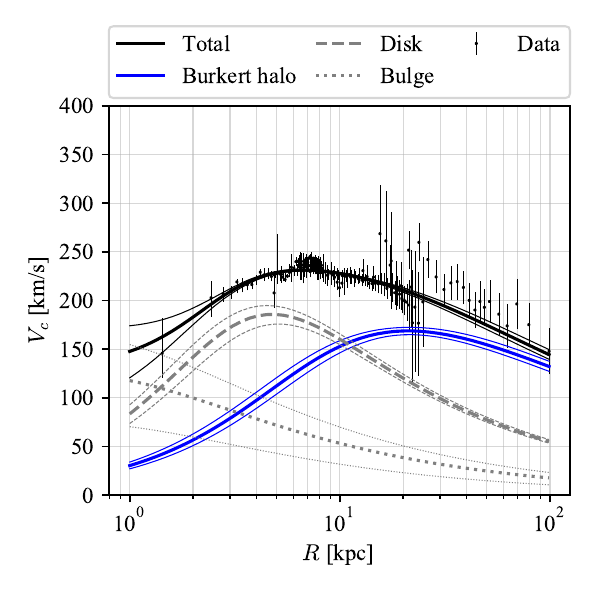}
		\includegraphics[width=0.49\textwidth]{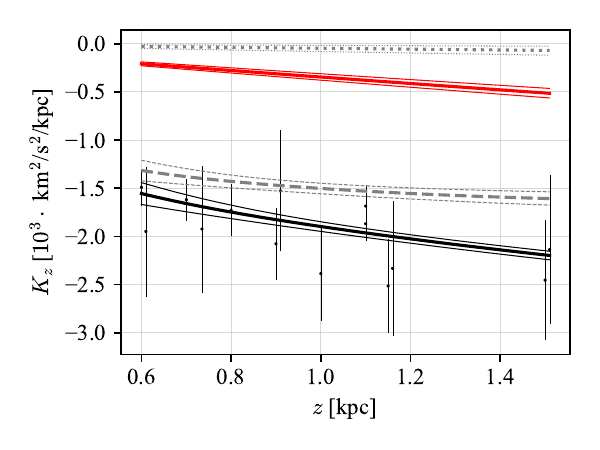}
		\includegraphics[width=0.49\textwidth]{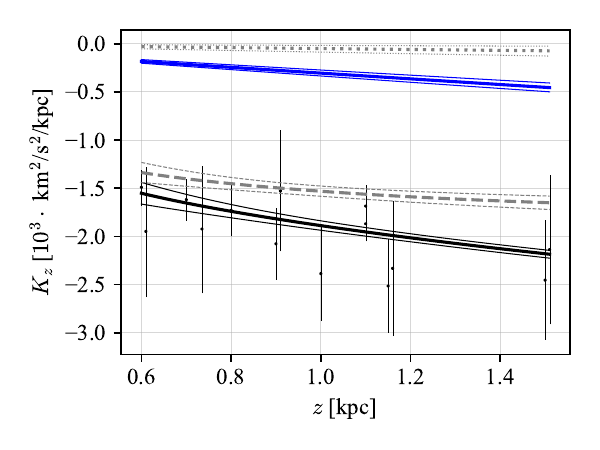}
		\includegraphics[width=0.49\textwidth]{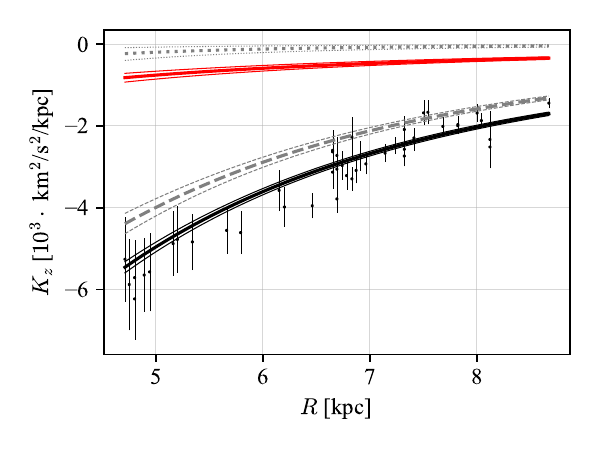}
		\includegraphics[width=0.49\textwidth]{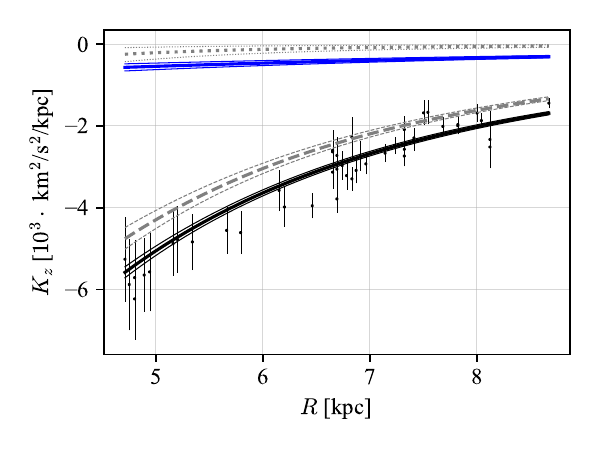}
		\caption{Comparison of the observational data and the results obtained from MCMC sampling for the galactic circular velocity and vertical force. The plots on left (right) hand side correspond to the results obtained under the assumption of NFW (Burkert) density profile. The thick lines correspond to median values while the thin lines denote the 68\% h.p.d. bands.}
		\label{fig:mw_fits}
	\end{figure}
	
	\begin{figure}[p]
		\includegraphics[width=\textwidth]{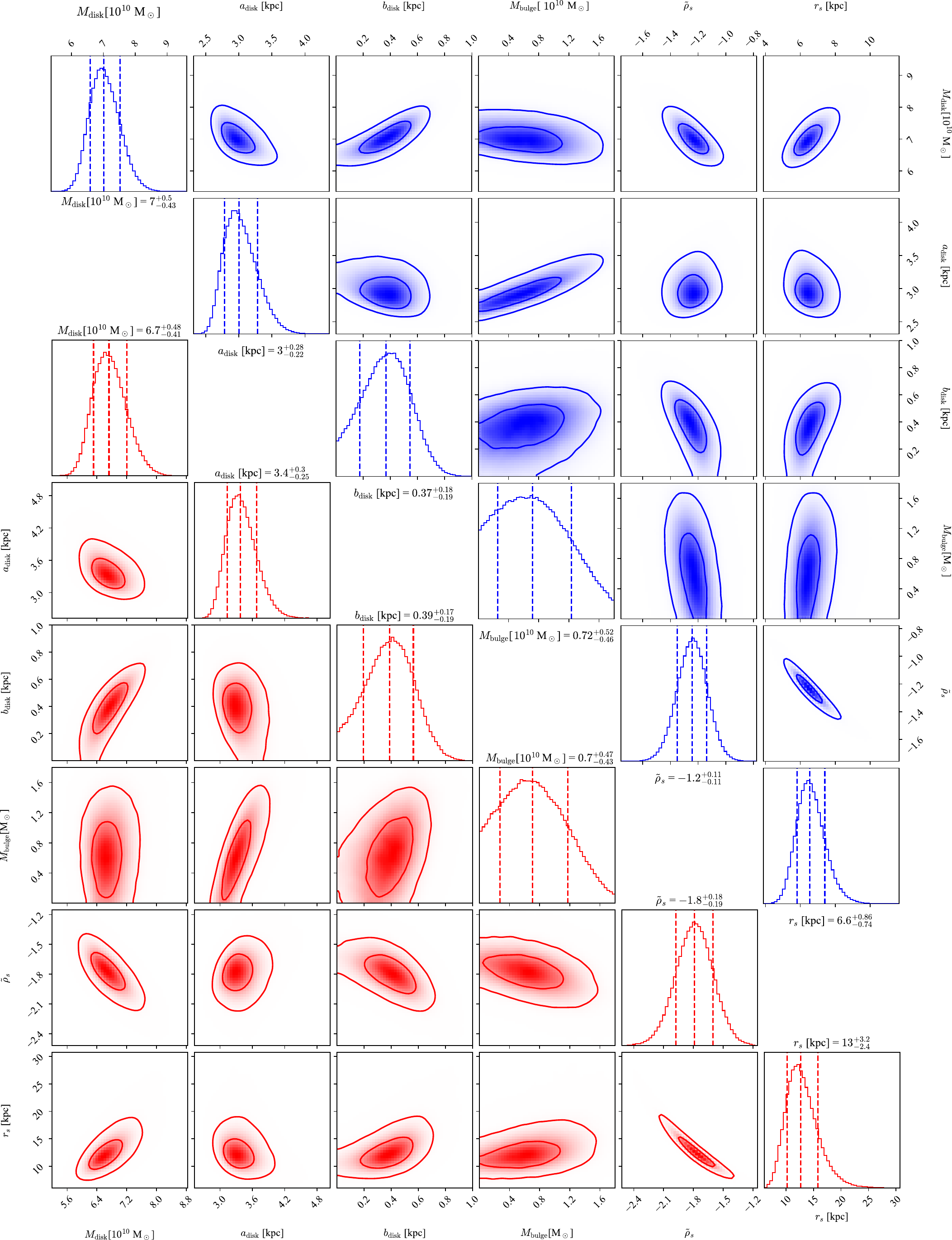}
		\caption{Single parameter posterior distributions and their pairwise correlations as obtained in the MCMC sampling for NFW (red) and Burkert (blue) DM density profile. The vertical lines in the posteriors mark median as well the 68\% h.p.d. interval, while the contours correspond to 39\% and 63\% h.p.d. regions.}
		\label{fig:mw_corner}
	\end{figure}
	
	In this section, we present our findings regarding the aforementioned parametric models of the galaxy which we fit to the kinematic data. The MCMC sampling showed good convergence and resulted in reduced chi-squared~\footnote{As an approximate measure for the goodness of the fit we use the reduced chi-squared, i.e. $\chi^2_\s{red} = \frac{\chi^2_\s{min}}{N - P}$, where $\chi^2_\s{min}$ is the chi-squared computed with asymmetric errors (analogously to the way it was done in the likelihood function) for the model parameters with the highest probability density, while $N$ and $P$ are the numbers of data points and model parameters used in the analysis.} of $\chi^2_\s{red} = 0.42$ and $\chi^2_\s{red} = 0.68$ under the assumption of NFW and Burkert density profile respectively. The obtained values of $\chi^2_\s{red}$ indicate that the adopted parametric models are flexible enough to accommodate the data and, at the same time, sufficiently simple to avoid over-fitting. This can also be nicely seen from the plots in figure~\ref{fig:mw_fits} where we show the over-all consistency of the obtained rotation curves and vertical force profiles with the observations, as well as the contributions of individual galactic components. By comparing the resulting rotation curves for models with NFW and Burkert DM density profiles, one can see that the outermost rotation curve measurements lead to a minor preference for the NFW profile; these are better described by more extended DM halos while the galactic mass decomposition generically leads to Burkert halos with smaller $r_s$ in comparison with the NFW case. However, we suspect that the precise values of $\chi^2_\s{red}$ might change by adopting a different parametrization of the baryonic disc (e.g. using the double-exponential ansatz~\eqref{eqn:exp_disc} or multiple disc components) and, therefore, we restrain ourselves from interpreting our results as substantial evidence in favor of the NFW density profile. The plots of rotation curves also clearly show the effect of degeneracy between individual galactic components since their 68\% highest probability density (h.p.d.) bands are significantly broader than the 68\% h.p.d. of the corresponding total. On the other hand, from plots in the middle and lower panels of figure~\ref{fig:mw_fits} one can see that the vertical force measurements impose significant constraints only on the disc component, as the contributions of DM and bulge to $K_z(R,z)$ are significantly smaller. Therefore, as we anticipated, the vertical force data provide valuable complementary information that helps in breaking the degeneracy between DM halo and baryonic disc while the over-all fit is primarily driven by the rotation curve measurements.
	
\begin{figure}[h]
		\centering
		\includegraphics[width=5in]{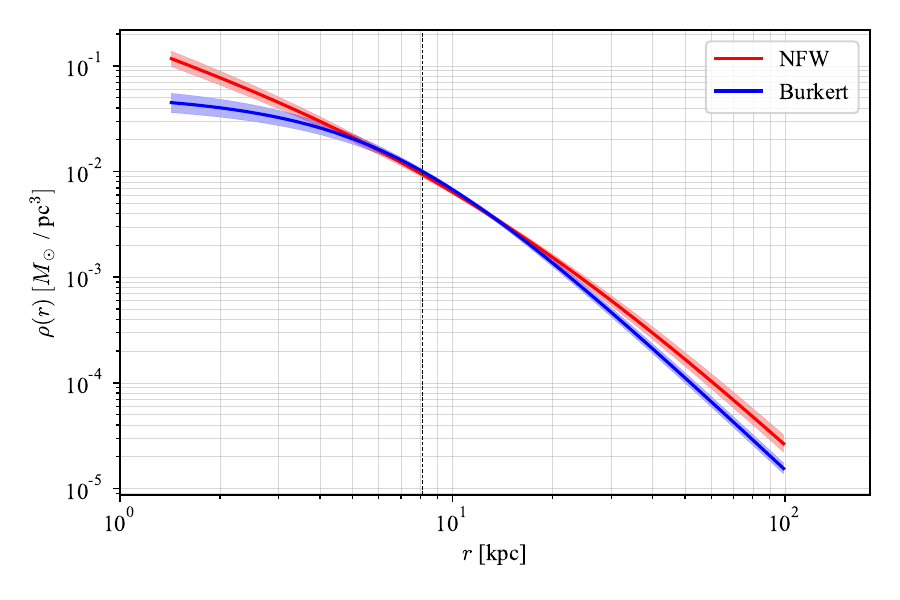}
		\caption{The inferred DM densities as a function of galactocentric distance with their corresponding 68\% h.p.d. bands. The vertical dashed line marks the solar galactocentric distance $R_\odot$.}
		\label{fig:mw_rho}
	\end{figure}

	In figure~\ref{fig:mw_corner}, we present the posterior distributions of the parameters as well as their pairwise correlations for the sampled galactic mass models. The obtained mass of the bulge is slightly smaller than the typical values found throughout the literature~\cite{sofue_rotation_2013,mcmillan_mass_2017}. However, this is most likely a consequence of the fact that the MN parameterization leads to a higher density of the baryonic disc at the center, while most other studies assumed an exponential disc and therefore attributed a larger mass to the bulge component. By looking at the total mass within the inner two kpc, the obtained results are in good agreement with the microlensing constraints on the optical depth towards the galactic center~\cite{portail_made--measure_2015,wegg_moa-ii_2016}. In contrast with the galactic bulge, the inferred mass of the baryonic disc is slightly larger than the sum of exponential stellar and gaseous discs typically found in the literature. This is partly due to the aforementioned effect, i.e. attributing a larger fraction of the central mass to the disc instead of the bulge, and partly due to the slower decline of the MN density distribution, which falls off in the radial direction only as $R^{-3}$ and not exponentially as the double-exponential ansatz, requiring larger total mass to explain the same dynamics at intermediate galactocentric radii. While this might seem problematic for the inferred DM density profiles, it plays no significant role as the majority of the ``excess" disc mass resides at radii where the baryonic contribution is strongly suppressed with respect to the DM counterpart. On the other hand, the inferred disc scale length, $a_\s{disc}$, and scale height, $b_\s{disc}$, agree well with the typical values found in the literature. Similarly, the results obtained for DM halo parameters fall within the range of values found in previous studies. The most significant difference with respect to the older works is perhaps a slightly smaller value of $r_s$ for both DM density profiles. For the local DM densities, we find the following median values and 68\% h.p.d. intervals:
	\begin{align}
	\label{eqn:local_density_NFW}
	\rho_\st{NFW}(R_\odot) & = 0.941^{+0.053}_{-0.057} \cdot 10^{-2} \; \s{M}_\odot / \s{pc}^3 = 0.357^{+0.020}_{-0.021} \; \s{GeV} / \s{cm}^3 \; , \\
	\label{eqn:local_density_BUR}
	\rho_\st{BUR}(R_\odot) & = 1.00^{+0.054}_{-0.057} \cdot 10^{-2} \; \s{M}_\odot / \s{pc}^3 \; = 0.381^{+0.020}_{-0.022} \; \s{GeV} / \s{cm}^3 \; .
	\end{align}
	The above values agree well with a number of recent studies~\cite{mcmillan_mass_2017,salas_estimation_2019,sofue_rotation_2020,ablimit_rotation_2020}
	On the other hand, several older works showed preference for somewhat larger values of $\rho(R_\odot)$ -- see, e.g.,~\cite{salucci_dark_2010,catena_novel_2010,nesti_dark_2013,sofue_rotation_2013,iocco_evidence_2015}. This is most probably a consequence of the updated rotation curve measurements since these tend towards lower and faster declining $V_c(R)$ at the solar radius with respect to previous determinations. The robustness of our results around $R_\odot$ is further supported by the fact that the derived values of $\rho(R_\odot)$ for the two DM density profiles are consistent with each other despite noticeable differences in the associated baryonic components.  In figure~\ref{fig:mw_rho}, we present the obtained DM density profiles as a function of galactocentric distance together with their 68\% h.p.d. bands. From there one can see that they are in good agreement over a radial range between $5 \; \s{kpc}$ and $20 \; \s{kpc}$, which coincides with the radii at which rotation curve measurements are the most accurate. On the other hand, outside this range, the Burkert profile leads to DM densities that are lower by a factor of $\sim 2$ with respect to the values at minimal and maximal radius spanned by the rotation curve measurements used in our work. This discrepancy can also be seen in the inferred values of the virial mass (which we define as the DM mass enclosed within a sphere of radius $r_{200}$ chosen such that its average density is equivalent to 200 times the cosmological critical density):
	\begin{align}
	M_\st{NFW}(r_{200}) = 7.7^{+1.4}_{-1.1} \cdot 10^{11} \; \s{M}_\odot \; , \\
	M_\st{BUR}(r_{200}) = 5.1^{+0.5}_{-0.5} \cdot 10^{11} \; \s{M}_\odot \; .
	\end{align}
	The above values are, however, somewhat lower then the ones obtained in other recent works~\cite{collaboration_gaia_GC_dsph_2018,posti_mass_2019,watkins_evidence_2019,callingham_mass_2019,eadie_cumulative_2019,vasiliev_proper_2019,karukes_robust_2020,wang_mass_2020,cautun_milky_2020,fritz_mass_2020}. The reason for this is most probably the choice of kinematic measurement -- in our work we have adopted a compilation of fairly direct $V_c(R)$ and $K_z(R,z)$ determinations, while the aforementioned virial mass estimates mostly rely on dynamical models for the Milky Way's globular clusters and satellite galaxies, which come with many underlying assumptions. On the other hand, we explicitly checked that by including the virial mass estimate of~\cite{collaboration_gaia_GC_dsph_2018} as an additional constraint in our fits does not lead to significant differences in the inferred local DM density or baryonic parameters that are crucial for accurately determining the local phase-space distribution of DM. While it can have a noticeable impact on the DM density at the outskirts and innermost part of our galaxy, it should be noted that the inclusion of $M_\s{vir}$ as an additional independent measurement comes at the cost of significantly increasing the total $\chi^2_\s{min}$, namely by 5 for NFW and 11 for Burkert density profile.
	
	In conclusion, our results provide conservative bounds on the NFW and Burkert DM density profiles as we restrained from making strong assumptions regarding the baryonic components.
	The inferred DM and baryonic parameters show fair agreement with previous studies. However, some caution is advised regarding the disc and bulge mass as they are often parameterized by different functional forms than the ones used in this work. On the other hand, in our analysis, we complemented the newest determinations of circular velocities by independent constraints regarding the galactic vertical force which helped us to disentangle the baryonic and DM components, but also provided more substantial leverage on constraining the gravitational potential of the galactic disc. This is particularly important for our analysis since the flattened baryonic component can have intricate consequences on the velocity distribution of the surrounding DM particles.
	
	\section{Implications for direct detection}
	\label{sec:direct_detection}
	
	Direct detection (DD) experiments provide a unique probe for investigating possible interactions between DM and baryons. By using large exposures of selected target materials they are capable of setting strong limits on the scattering rate of atomic nuclei with the galactic DM particles. For a DM candidate with a given differential DM-nucleus cross-section, $\s{d}\sigma / \s{d}E_r$, the expected differential recoil rate per target nucleus can be computed as~\cite{goodman_detectability_1985}:
	\begin{align}
	\frac{\mathrm{d}R}{\mathrm{d}E_r} = \frac{1}{m_A m_\chi} \cdot \int_{|\vec{v}| > v_\textrm{min}} \s{d}^3v \; f(\vec{x}, \vec{v}) \cdot v \cdot \frac{\s{d} \sigma}{\s{d} E_r} \label{eqn:rate_dif} \\
	\textrm{with} \;\;\; v_{\textrm{min}} = \sqrt{\frac{m_A E_r}{2 \mu^2_{A \chi}}} \; \; \; , \; \; \; \mu_{A \chi} = \frac{m_A m_\chi}{m_A + m_\chi} \; , \nonumber
	\end{align}
	where $E_r$ is the recoil energy, $m_{A/\chi}$ the nucleus/DM mass and $v = |\vec{v}_\st{LAB}|$ the velocity of DM particle in the detector's (LAB) frame. The latter is a sum of the DM velocity in the galactic rest frame, $\vec{v}_0$, the velocity of the local standard of rest (LSR), $\vec{v}_\st{LSR}$, the peculiar motion of the Sun, $\vec{v}_\odot$, and the Earth's circular velocity around the Sun, $\vec{v}_\s{circ}$:
	\begin{align}
	\label{eqn:v_lab}
	\vec{v}_\st{LAB} = \vec{v}_0 + \vec{v}_\st{LSR} + \vec{v}_\odot + \vec{v}_\s{circ} \;.
	\end{align}
	For the LSR and peculiar motion of the Sun we adopt values consistent with the kinematic data used in section~\ref{sec:rotation_curve}, namely $\vec{v}_\s{LSR} = (0, 231, 0) \; \s{km/s}$ and $\vec{v}_\odot = (11.1, 12.24, 7.25) \; \s{km/s}$, while for Earth's circular motion we use $|\vec{v}_\s{circ}| = 30 \; \s{km/s}$ along with orbital parameters reported in~\cite{freese_colloquium_2013}. Since the orientation of $\vec{v}_\s{circ}$ varies throughout the year, which leads to annual modulation of the signal that we discuss in grater detail in section~\ref{sec:direct_detection_modulation}, all the quantities obtained in the following are given as yearly averages unless stated otherwise. For spin-independent (SI) interactions, where DM is typically assumed to scatter coherently with all the nucleons, the differential cross-section can be expressed as:
	\begin{align}
	\frac{\s{d} \sigma}{\s{d} E_r} = \frac{m_A \sigma^{\textrm{SI}}_{n}}{2 \mu^2_{A \chi} v^2} A^2 F^2(E_r) \; ,
	\end{align}
	where $\sigma^{\textrm{SI}}_{n}$ is the SI DM-nucleon cross-section at zero momentum transfer, $A$ the mass number of the target nucleus and $F(E_r)$ the nuclear form factor. As can been seen from the above expression, the SI differential cross-section introduces an additional factor of $v^{-2}$ within the integral of equation~\eqref{eqn:rate_dif}, which also appears in the case of spin-dependent (SD) interactions, however, this is not always true for more general scattering operators. In either case, on can factorize equation~\eqref{eqn:rate_dif} into a term determined by the specific particle physics model under consideration and an astrophysical factor which is a convolution of the process's velocity dependence with the DM distribution function. For SI and SD case the relevant integral takes the following form~\footnote{Note that the definition of $g(v_\textrm{min})$ in equation~\eqref{eqn:g}, as well as $h(v_\textrm{min})$ in equation~\eqref{eqn:h}, differs from the one typically found in the literature by a factor of $\rho_\odot$. This is because in our approach the DM density and its velocity distribution are simultaneously inferred from the kinematic observations on the level of phase-space distribution.}:
	\begin{align}
	\label{eqn:g}
	g(v_\textrm{min}) \equiv  \int_{|\vec{v}| > v_\s{min}} \s{d}^3v \; \frac{f(\vec{x}, \vec{v})}{v} \; .
	\end{align}
	However, it is often desired to go beyond the simplest scattering operators since there are many other ways in which DM can couple to the nucleons. In order to address the wide range of possibilities, a fully general set of non-relativistic effective scattering operators has been assembled -- for their systematic treatment see~\cite{fitzpatrick_effective_2013,anand_model-independent_2013,dent_general_2015,bishara_chiral_2017}. For many phenomenologically interesting models the leading order contribution to the differential cross-section can also be velocity independent, hence, it is useful to additionally define:
	\begin{align}
	\label{eqn:h}
	h(v_\textrm{min}) \equiv  \int_{|\vec{v}| > v_\s{min}} \s{d}^3v \;\; f(\vec{x}, \vec{v}) \cdot v \; .
	\end{align}
	
	It turns out that the above functions, $g(v_\s{min})$ and $h(v_\s{min})$, cover the velocity dependencies of all possible non-relativistic effective scattering operators expanded up to the quadratic order in momentum transfer and relative velocity. Therefore, their accurate determination is of great importance for understanding the direct detection constraints on DM-nucleus interactions. While the SHM predicts the total event rate reasonably well in the regime where $v_\s{min} \ll v_\s{esc}$, it becomes increasingly unrealistic as $v_\s{min}$ approaches the sharp cut-off at $v_\s{esc}$. As a consequence, the limits on DM-nucleon cross-section can be significantly over-estimated at the lower end of the DM mass range probed in a given detector. It can also significantly impact the predicted energy spectrum of the nuclear recoils, which plays an important role in rejecting backgrounds but could also help in identifying the structure of the coupling upon successful detection of DM. To obtain more accurate predictions for these quantities, one has to adopt a more physical approach of constructing the DM distribution function, such as the Eddington's inversion method, which was extensively studied in the context of DM searches throughout the last decades~\cite{vergados_new_2003,catena_local_2012,lacroix_anatomy_2018,lacroix_predicting_2020}.
	However, the axisymmetric modeling that we advocate in our work has even further advantages over the Eddington-like approaches that are based on spherical symmetry. Most importantly, it allows us to compute the velocity anisotropy of the DM particles that arises due to the presence of the baryonic disc and can have a sizable effect on several observables. For example, it affects the expected amplitude of the annual modulation in the recoil rate, which arises due to the time-dependent mapping of $f$ from the galactic to the LAB rest frame in equation~\eqref{eqn:rate_dif}. Since this is a unique feature expected only for DM-induced signals, its accurate modeling plays an essential role in rejecting various backgrounds. The amplitude of modulation is equally sensitive to the rotational properties of the DM halo, which can again be self-consistently included within the axisymmetric model. Furthermore, spherically symmetric models can lead to a less accurate reconstruction of the DM mass and cross-section as well as a poorer determination of the DM-nucleon interaction type in near-future experiments if DM-induced signals are detected. Finally, the velocity anisotropy and rotation of the DM halo can also strongly affect the expected DM signal in directional DD experiments, which are capable of reconstructing the direction of observed nuclear recoils. While such experiments are currently still inferior to the classical noble gas and crystalline detectors, they are expected to become crucial in the future when the sensitivities reach the neutrino floor, and directional reconstruction of events will be essentially the only way of distinguishing genuine DM signals from neutrino-induced backgrounds~\cite{monroe_neutrino_2007,vergados_can_2008,strigari_neutrino_2009,billard_implication_2014,grothaus_directional_2014,mayet_review_2016,ohare_can_2020,vahsen_cygnus_2020}.
	
	In the following we first present our results regarding the uncertainties in $g(v_\s{min})$ and $h(v_\s{min})$ on a set of benchmark models, assuming either NFW or Burkert halo profile with and without rotation. Subsequently, we take a closer look at the impact of our findings on the interpretation of direct searches. First, we compare the inferred cross-section limits based on the considered halo models, paying particular attention to the currently most stringent XENON bounds. After that, we turn our attention to various aspects of direct searches that would become crucial upon successful detection of possible DM signals. In this context we first demonstrate the importance of carefully constraining the astrophysical factors for accurate reconstruction of DM properties in near-future experiments, secondly, we examine the impact of detailed phase-space modeling on the expected annual modulation of the nuclear recoil rate and, finally, we analyze the effect of axisymmetric models on the expected directional distribution of nuclear recoils.
	
	\subsection{Bracketing of astrophysical factors}
	\label{sec:astrophysical_factors}
	
\begin{figure}[h]
		\includegraphics[width=0.49\textwidth]{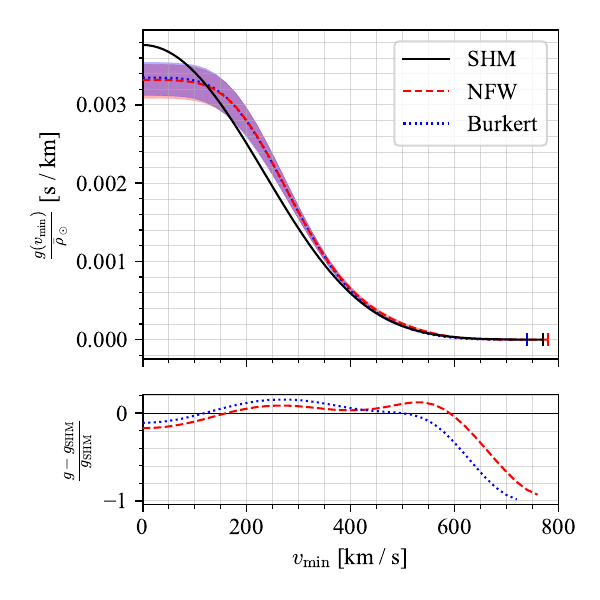}
		\includegraphics[width=0.49\textwidth]{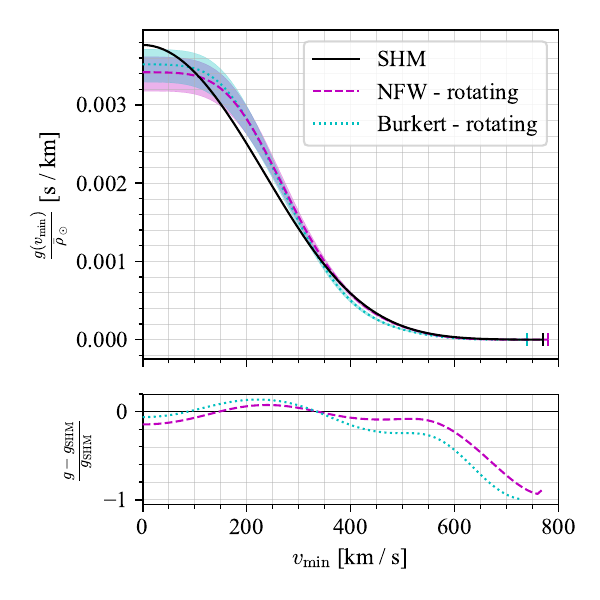}
		\includegraphics[width=0.49\textwidth]{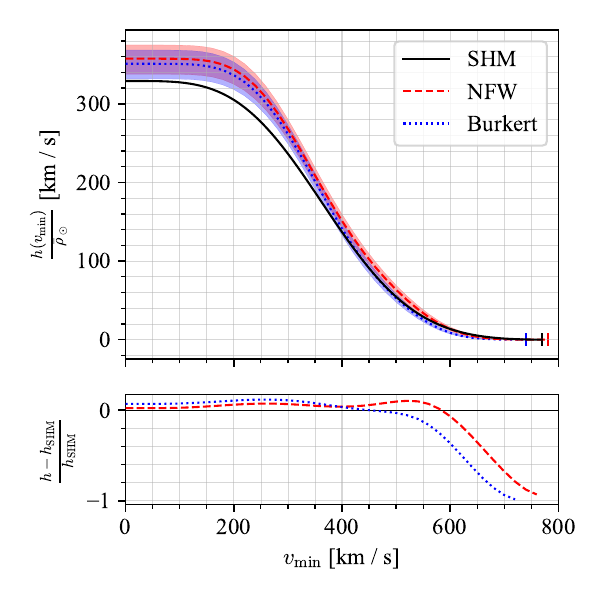}
		\includegraphics[width=0.49\textwidth]{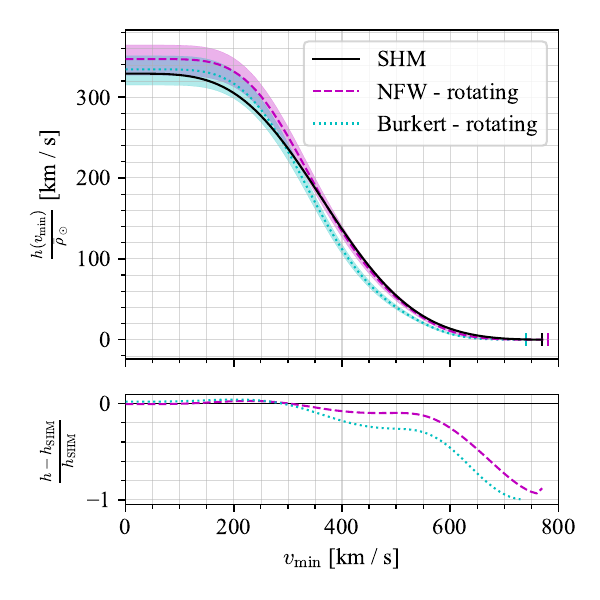}
		\caption{Astrophysical factor that enter direct detection experiments as a function of the minimum scattering velocity for the two relevant velocity dependences of the cross-section (defined in equations~\eqref{eqn:g} and \eqref{eqn:h}), normalized by the corresponding median local DM density. Solid lines show the median value, while the bands correspond to 68\% credibility regions. The results are shown for NFW and Burkert density profiles, assuming non-rotating halos, as well as rotating ones with spin parameter $\lambda = 0.04$ and azimuthal velocity profile given by equation~\eqref{eqn:rotation_profile}.}
		\label{fig:mw_astro_fact}
	\end{figure}

	One of the key results of this work are the refined predictions for the astrophysical factors that crucially enter the interpretation of DD experiments. By computing the astrophysical factors for $10^4$ randomly picked samples from the MCMC scan described in section~\ref{sec:mass_depomposition} we obtained their posterior distributions which in turn allowed us to extract their central values and credibility regions. In figure~\ref{fig:mw_astro_fact} we present our results for $g(v_\s{min})$ and $h(v_\s{min})$ under different assumptions regarding the DM density profile (NFW or Burkert) and halo rotation (non-rotating or co-rotating with $\lambda = 0.04$ as described in section~\ref{sec:DM_assumptions}). They are normalized by the corresponding average densities, reported in equations~\eqref{eqn:local_density_NFW} and~\eqref{eqn:local_density_BUR}, for the purpose of easier comparison with previous works where the DM density was usually treated as an independent quantity. In the same figure we additionally show the predictions of the SHM, assuming $\sigma = 156 \; \s{km/s}$ and $v_\s{esc} = 544 \; \s{km/s}$.~\footnote{The given values of SHM parameters closely match the typical values used throughout the literature, see, e.g.,~\cite{green_astrophysical_2017,aprile_first_2017,supercdms_collaboration_low-mass_2018,the_darkside_collaboration_low-mass_2018,akerib_results_2017}. However, they are slightly different because the values used in this work were adjusted to the average values obtained from the highest probability density parameters of mass models under the assumption of NFW and Burkert density profiles.} From the plots one can see that the SHM mostly falls within the 68\% highest probability density (h.p.d.) bands spanned by the equilibrium models, however, larger deviations can occur. In particular, for $g(v_\s{min})$ the SHM yields up to 20\% larger values at $v_\s{min} \lesssim 50 \, \s{km/s}$ and 20\% lower values at intermediate $v_\s{min} \sim 250 \, \s{km/s}$, which both fall outside of the 68\% h.p.d. bands for all the considered equilibrium models. Further differences can also be observed in the shape of $g(v_\s{min})$ and $h(v_\s{min})$, especially when comparing the SHM with median values of the equilibrium models. Most notably, at low $v_\s{min}$ the SHM predicts a much faster decline of both astrophysical factors. This arises because the equilibrium models exhibit smaller velocity dispersion of the DM particles along the azimuthal direction and larger velocity dispersion in the meridional ($\hat{R}$-$\hat{z}$) plane which in turn strongly suppresses the probability for scatterings with $v \lesssim 100 \; \s{km/s}$ in the LAB frame. Further important differences in the astrophysical factors are present at large $v_\s{min}$. While there the absolute values of $g(v_\s{min})$ and $h(v_\s{min})$ are small, the relative difference between astrophysical factors computed through equilibrium models and the SHM becomes very large as can be seen from their ratios, plotted in the lower panels of figure~\ref{fig:mw_astro_fact}. Due to this reason the SHM is known to over-constrain the DM-nucleon cross-section at low DM masses while our approach provides much more conservative bounds. On the other hand, the equilibrium models lead to appreciably more consistent values of the astrophysical factors with substantial overlap of their 68\% h.p.d. regions for all the considered cases. At low $v_\s{min}$ slightly larger values of $g(v_\s{min})$ ($h(v_\s{min})$) are predicted by the halo model with Burkert (NFW) profile  while at $v_\s{min} \gtrsim 350 \; \s{km/s}$ the NFW profile predicts larger values for both functions and at the same time also comes with higher escape velocity. By comparing the non-rotating halos with the rotating ones we can see that the change in $g(v_\s{min})$ and $h(v_\s{min})$ is again relatively small. The key difference arises from the fact that co-rotating halos lead to larger number of scatterings at low LAB velocities which results in slightly larger $g(v_\s{min})$ and smaller $h(v_\s{min})$. For the same reason the astrophysical factors of rotating halos are even further suppressed at large $v_\s{min}$ which can have, as already discussed above, significant consequences for the interpretation of DD searches at small DM masses. For the convenience of future analyses we provide the tabulated values and associated credibility intervals of the obtained astrophysical factors.
	
	\subsection{Implications for direct detection experiments}
	
	\subsubsection{Cross-section bounds}
	
\begin{figure}[h]
		\includegraphics[width=0.49\textwidth]{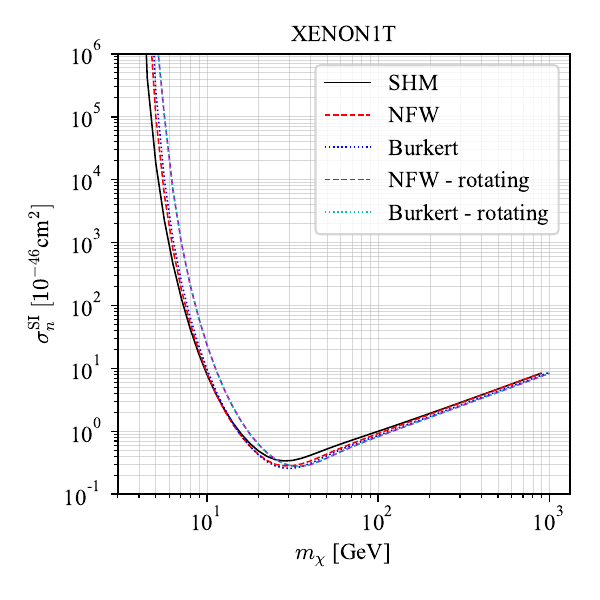}
		\includegraphics[width=0.49\textwidth]{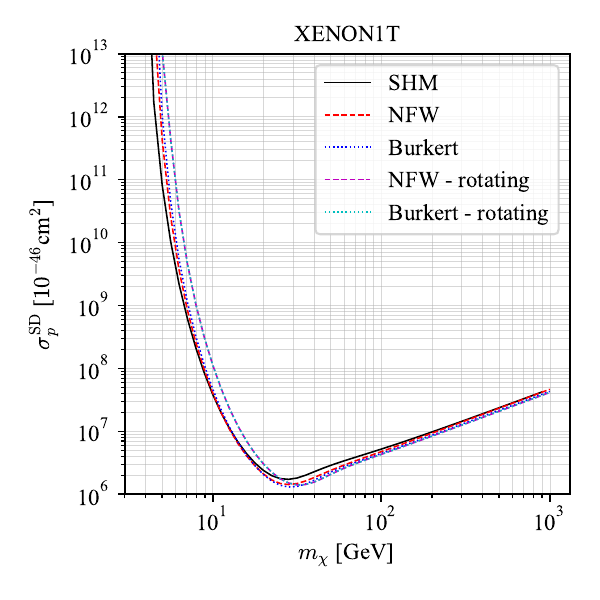}
		\caption{The XENON1T upper limits on SI (left) and SD (right) cross-section, as obtained for the SHM~\cite{aprile_dark_2018} and non-/co-rotating axisymmetric models fit to the kinematic data using NFW and Burkert density profile. For the latter, exclusion limits corresponding to the lower 95\% credibility bound on $g(v_\s{min})$ are shown.}
		\label{fig:mw_sigma}
	\end{figure}

	Using the tabulated values of $g(v_\s{min})$ and $h(v_\s{min})$ one can easily obtain the limits on arbitrary DM-nucleon coupling through the use of existing software, such as DDCalc~\cite{bringmann_darkbit:_2017,the_gambit_collaboration_global_2019}. In figure~\ref{fig:mw_sigma}, we present the upper limits on the DM-nucleon cross-section (assuming either SI interactions with an equal coupling of DM to neutrons and protons or SD interactions with DM coupling exclusively to neutrons) obtained from XENON1T null results for the benchmark halo models and the SHM that was used in the original interpretation of data by XENON collaboration~\cite{aprile_dark_2018}. As can be seen from the plots, there can be significant deviations, especially at low DM masses, where the bounds derived from SHM are over-constraining due to the thick tail of the truncated Maxwell-Boltzmann velocity distribution. By using the lower 95\% confidence values of $g(v_\s{min})$, the bounds on cross-section at $m_\chi \lesssim 5$ GeV are relaxed by a factor of 5 and 15 for non-rotating NFW and Burkert halos, respectively, and almost two orders of magnitude under the assumption of a co-rotating NFW halo. The differences become smaller at larger DM masses, however, still significant changes of the bounds are present at DM masses near to the peak sensitivity of the detector, i.e. at $m_\chi \sim 30$ GeV. In contrast to the low $m_\chi$ range, there the SHM leads to roughly 30\% weaker limits than the considered equilibrium models. For even large DM masses the difference becomes less than 10\%, with SHM resulting in the least stringent bounds. As it was already observed in the previous section on the level of astrophysical factors, the differences among the axisymmetric halo models are appreciably smaller and significant deviations occur only at small $m_\chi$, corresponding to large $v_\s{min}$. The DM distribution function with NFW density profile typically leads to few per cent weaker limits at large DM masses, however, the bounds can become significantly stronger at $m_\chi < 30$ GeV with respect to the ones associated with Burkert halos, which is mainly due to the larger escape velocity of the NFW model. Similarly, non-rotating halos tend to slightly relax the limits at large DM masses while they become considerably more constraining for small $m_\chi$, especially in combination with the NFW density profile.
	In figure~\ref{fig:mw_sigma}, we explicitly recompute the limits only for the XENON1T experiment. However, due to the universality of $g(v_\s{min})$, a similar behavior is expected also in other DD experiments. In deriving the cross-section limits, we also restricted our attention only to the standard SI and SD interactions but based on the astrophysical factors shown in figure~\ref{fig:mw_astro_fact} similar general trends are expected for any other type of scattering operator between DM and baryons.
	
	\subsubsection{Reconstruction of DM properties in next generation experiments}
	
	Further motivation for accurate determination of the astrophysical factors comes from the next-generation detectors which are currently under construction, such as DARWIN~\cite{aalbers_DARWIN_2016} and DarkSide-20k~\cite{aalseth_darkside-20k_2018}. By allowing for exploration of cross-sections that are orders of magnitude below the currently most stringent bounds, they provide an opportunity to undisputedly detect DM-nucleus interactions in previously un-probed regions of the parameter space. In this context, an important question arises: how well will these detectors be able to characterize the properties of DM if they detect a signal appreciably larger then the expected backgrounds? To demonstrate the importance of astrophysical factors in such inquiries, we perform two different types of benchmarks based on simulated datasets and expected characteristics of the DARWIN and DarkSide-20k detectors. First, we study the prospects of pinning-down the mass and cross-section of DM for a fixed scattering operator and, secondly, the ability to discriminate among different scattering operators when using the information from individual or both aforementioned experiments.
	
	Successful detection of DM-induced nuclear recoils would open a unique window for studying the properties of DM particles. However, it has been noted in numerous previous works (see, e.g.~\cite{green_determining_2008,akrami_how_2011,strege_fundamental_2012,catena_analysis_2014,edwards_dark_2018}) that accurate determination of the DM mass and the relevant coupling parameter can still be difficult due to degeneracies in the expected signals which are further worsened by limited precision of observations and uncertainties related to the local DM velocity distribution. As we demonstrate in the following, using refined phase-space models for the galactic DM has significant advantages in the reconstruction of DM properties in next-generation experiments. We consider two possible DM-nucleus interaction types, namely the standard SI interaction and coupling through the anapole moment. The latter is generated by the following relativistic operator:
	\begin{align}
	\mathscr{L} \supset \mathcal{A} \, \bar{\chi} \gamma^\mu \gamma^5 \chi \partial^\nu F_{\mu \nu} \; ,
	\end{align}
	where $\mathcal{A}$ is the effective dimensionful coupling constant, $\chi$ denotes the DM field and $F_{\mu \nu}$ the electromagnetic field tensor. We chose to study the possibility of interactions through anapole moment due to its peculiar structure in the non-relativistic limit, as it depends on both astrophysical factors, $g(v_\s{min})$ and $h(v_\s{min})$ -- for the explicit expression of the corresponding differential cross-section and it's derivation see, e.g.,~\cite{fitzpatrick_dark_2010,del_nobile_direct_2014}.
	
	In order to compare the prospects of reconstructing the DM mass and coupling strength under different assumptions regarding the DM distribution, one first has to obtain the average energy-binned DM signals that are expected for the chosen benchmark models in a given detector. To achieve this, some fiducial astrophysical factors have to be assumed. In this work we chose to adopt $g(v_\s{min})$ and $h(v_\s{min})$ obtained from the axisymmetric PSDF corresponding to the h.p.d. galactic parameters for NFW halo derived in section~\ref{sec:mass_decomposition_results} (note that these differ from the median values plotted in figure~\ref{fig:mw_astro_fact}, but do fall within the corresponding 68\% credibility band). Furthermore, to compute the expected signals one has to additionally specify the DM mass, for which we consider two representative values, $m_\chi = 30 \, \s{GeV}$ and $m_\chi = 100 \, \s{GeV}$, and the DM-nucleus cross-section, which we fixed slightly below the current most stringent limits for a given $m_\chi$ and interaction type. To simulate real observations, we created 1000 stochastic realizations of energy-binned mock datasets for each of the detectors with their appropriate energy resolutions. Each energy bin was populated with a number of events drawn from a Poisson distribution, whose mean rate corresponds to the sum of the expected number of signal and background events in a given energy bin, which were both computed using the DDCalc software~\cite{bringmann_darkbit:_2017,the_gambit_collaboration_global_2019}. Subsequently, we performed Bayesian sampling on each of the 1000 stochastic datasets, which upon combining the obtained samples allowed us to include the uncertainties related to the Poissonian noise that is inherently present in the reconstruction of DM parameters from real experimental data. For each dataset, $\mathcal{D}$, we obtained the posterior distribution of the DM mass and coupling parameter according to the following Poissonian likelihood function:
	\begin{align}
	\label{eqn:dd_likelihood}
	\mathcal{L}(\mathcal{D}|\vec{\theta}) = \prod_{i=1}^{N_\s{bins}} \; \frac{n_i (\vec{\theta})^{\tilde{n}_{i}}}{\tilde{n}_i!} \cdot e^{-n_i(\vec{\theta})} \; ,
	\end{align}
	where $n_i(\vec{\theta})$ and $\tilde{n}_i$ are the predicted and observed number of events in $i$th energy bin, respectively, while $\vec{\theta}$ is the vector of model parameters, namely the DM mass and cross-section. The above likelihood function can be easily extended to the situation where joined constraining power of multiple experiments is considered -- this can be achieved by letting $i$ run over all energy bins in all of the considered detectors. To additionally include the uncertainties stemming from the unknown underlying DM distribution, we chose to compute the astrophysical factors in each MCMC run for a random set of galactic parameters (but fixed parametric form of DM density profile and assumptions regarding the halo rotation) from the posterior distribution obtained in section~\ref{sec:mass_decomposition_results} which in this case served as a prior.~\footnote{By fixing the astrophysical factors throughout each of the MCMC runs we implicitly assume that the galactic parameters are constrained only by the astronomical observations discussed in section~\ref{sec:observations}. In principle, one could perform the sampling of the particle physics parameters and the galactic mass model simultaneously, however, such analysis would be significantly more challenging. Furthermore, it would most likely lead to very similar results since the constraints on the galactic parameters are primarily driven by the kinematic observations. On the other hand, our approximation leads to conservative results since the inferred credibility intervals regarding $m_\chi$ and the relevant cross-section must be broader than the ones which would be found in the more general approach.} On the other hand, for the DM mass and cross-section, we assumed flat logarithmic priors that spanned four orders of magnitude and were centered on the true values of the parameters. The sampling was performed using the affine invariant \textit{emcee} sampler~\cite{foreman-mackey_emcee_2013} with each run consisting of 50 walkers that evolved for 5000 steps. After completing the sampling, only the last 1000 steps of each walker were maintained to construct the joined posterior distribution for the DM mass and coupling parameter from all the completed runs.
	
	\begin{figure}[h]
		\centering
		\includegraphics[width=0.52\textwidth]{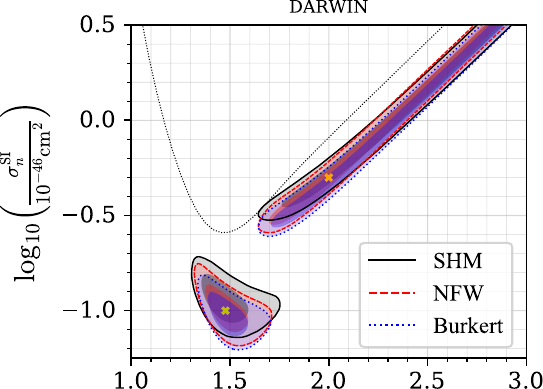}
		\includegraphics[width=0.465\textwidth]{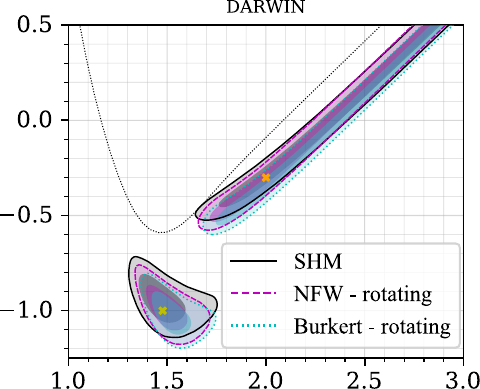}
		\vspace{0.5cm}
		\includegraphics[width=0.52\textwidth]{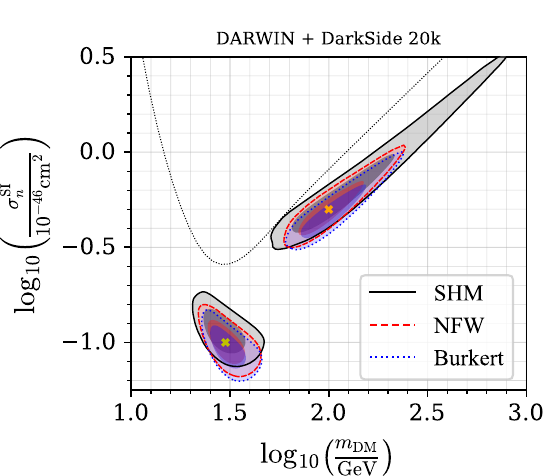}
		\includegraphics[width=0.465\textwidth]{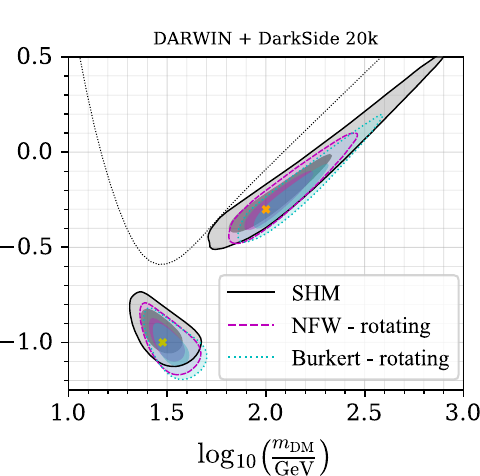}
		\caption{Reconstruction of the DM mass and $\sigma_n^\st{SI}$ upon successful detection of DM in the future experiments, namely DARWIN only in the top panel and DARWIN combined with DarkSide-20k in the bottom panel. The yellow and orange crosses mark the assumed true parameters that were used for generating the mock observations, \{$m_\chi = 30 \; \s{GeV}$, $\sigma_n^\st{SI} = 10^{-47} \s{cm}^2$\} and \{$m_\chi = 100 \; \s{GeV}$, $\sigma_n^\st{SI} = 5 \cdot 10^{-47} \s{cm}^2$\}, respectively, while the contours correspond to the 68\% and 95\% credibility regions obtained under different assumptions regarding the local DM distribution. The dotted line marks current most stringent limits for SI cross-section obtained by the XENON1T experiment.}
		\label{fig:dd_reconstruction_si}
	\end{figure}

	\begin{table}
		\centering
		\renewcommand{\arraystretch}{1.3}
		\begin{tabular}{c||c|c||c|c}
			Model & $\log_{10} \left( \frac{m_\chi}{\s{GeV}} \right)$ & $\log_{10} \left( \frac{\sigma_n^\st{SI}}{10^{-46}  \, \s{cm}^2} \right)$ & $\log_{10} \left( \frac{m_\chi}{\s{GeV}} \right)$ & $\log_{10} \left( \frac{\sigma_n^\st{SI}}{10^{-46}  \, \s{cm}^2} \right)$ \\ \hline \hline
			True & 1.48 & -1 & 2 & -0.3 \\ \hline
			SHM (DARWIN only) & $1.47^{+0.09}_{-0.07}$ & $-0.95^{+0.08}_{-0.07}$ & $2.85^{+0.78}_{-0.75}$ & $0.51^{+0.77}_{-0.71}$ \\
			NFW (DARWIN only) & $1.48^{+0.08}_{-0.07}$ & $-1^{+0.08}_{-0.07}$ & $2.75^{+0.84}_{-0.78}$ & $0.44^{+0.84}_{-0.76}$ \\
			Burkert (DARWIN only) & $1.49^{+0.08}_{-0.06}$ & $-1.03^{+0.07}_{-0.07}$ & $2.76^{+0.84}_{-0.77}$ & $0.41^{+0.84}_{-0.75}$ \\
			NFW - rotating (DARWIN only) & $1.5^{+0.08}_{-0.06}$ & $-0.99^{+0.08}_{-0.07}$ & $2.76^{+0.84}_{-0.77}$ & $0.44^{+0.84}_{-0.74}$ \\
			BUR - rotating (DARWIN only) & $1.52^{+0.08}_{-0.06}$ & $-1.02^{+0.07}_{-0.07}$ & $2.78^{+0.83}_{-0.76}$ & $0.42^{+0.82}_{-0.74}$ \\
			SHM (DARWIN + DarkSide-20k) & $1.47^{+0.08}_{-0.07}$ & $-0.95^{+0.08}_{-0.07}$ & $2.06^{+0.24}_{-0.15}$ & $-0.23^{+0.2}_{-0.11}$ \\
			NFW (DARWIN + DarkSide-20k) & $1.48^{+0.07}_{-0.06}$ & $-1^{+0.08}_{-0.07}$ & $2^{+0.13}_{-0.1}$ & $-0.3^{+0.11}_{-0.08}$ \\
			Burkert (DARWIN + DarkSide-20k) & $1.49^{+0.07}_{-0.06}$ & $-1.03^{+0.07}_{-0.07}$ & $2.02^{+0.13}_{-0.1}$ & $-0.32^{+0.11}_{-0.08}$ \\
			NFW - rotating (DARWIN + DarkSide-20k) & $1.5^{+0.07}_{-0.06}$ & $-1^{+0.07}_{-0.07}$ & $2.06^{+0.14}_{-0.11}$ & $-0.25^{+0.13}_{-0.09}$ \\
			Burkert - rotating (DARWIN + DarkSide-20k) & $1.52^{+0.07}_{-0.06}$ & $-1.02^{+0.07}_{-0.07}$ & $2.12^{+0.15}_{-0.11}$ & $-0.24^{+0.14}_{-0.1}$
		\end{tabular}
		\caption{True and reconstructed values of the DM mass and cross-section with the corresponding 68\% credibility intervals under different assumptions regarding the DM distribution function for the two benchmark scenarios with SI interaction.}
		\label{tab:dd_reconstruction_si}
	\end{table}
	
	The inferred 68\% and 95\% h.p.d. regions for SI interactions under the assumption of different DM distributions are shown in figure~\ref{fig:dd_reconstruction_si} while single-parameter credibility intervals are reported in table~\ref{tab:dd_reconstruction_si}, assuming $200 \, \s{t} \times \s{yr}$ and $100 \, \s{t} \times \s{yr}$ exposures for DARWIN and DarkSide-20k experiments, respectively.
	The sampling was performed using the posterior distribution of astrophysical factors obtained from (rotating and non-rotating) NFW and Burkert phase-space models, as well as the ones predicted by the SHM optimally adjusted to the fiducial NFW galactic mass model with appropriate Gaussian errors on its parameters (namely, $\rho_\odot = 0.941 \pm 0.021 \cdot 10^{-2} \, M_\odot / \s{pc}^3$, $\sigma = 157 \pm 4 \, \s{km/s}$ and $v_\s{esc} = 537 \pm 19 \, \s{km/s}$ while their correlations were neglected). As can be seen from the plots, all DM distribution models based on the axisymmetric PSDF provide better reconstruction (especially in the case of heavy DM) of the true parameters in comparison with the SHM, even though the latter was tuned to match the fiducial model. The SHM tends to over-predict the value of the coupling parameter while the inferred DM mass is also significantly biased, either towards lower values at $m_\chi = 30 \, \s{GeV}$ or higher values at $m_\chi = 100 \, \s{GeV}$. On the other hand, it is not surprising that the fits obtained under the assumption of NFW astrophysical factors always performed the best and lead to negligible (sub per cent) bias in the inferred central values, apart from the run with $100 \, \s{GeV}$ DM for the DARWIN detector only, where there is a severe degeneracy between the parameters. The parameter reconstruction under the assumption of Burkert astrophysical factors performed slightly worse, typically leading to a few per cent bias towards lower values of the coupling parameter, which is primarily due to the larger local DM density associated with this DM profile, while the reconstructed mass was found to lie very close to the true value. The reconstruction further worsened by performing the sampling using astrophysical factors associated with co-rotating halos, however, the obtained results were still better (or at least comparable to) the ones obtained under the assumption of SHM. The rotating models led to the inference of $\mathcal{O}(10\%)$ larger DM masses due to lower number of scatterings with velocities above the $v_\s{min}$ threshold, while the reconstructed values of cross-section remained largely unchanged.
	
\begin{figure}[h]
		\centering
		\includegraphics[width=0.52\textwidth]{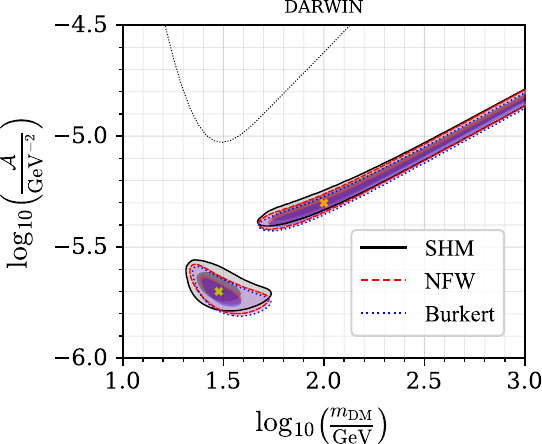}
		\includegraphics[width=0.47\textwidth]{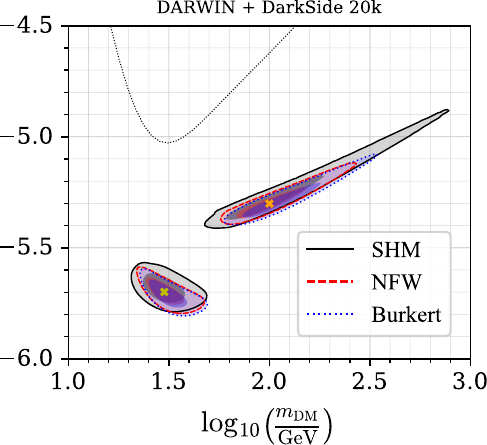}
		\vspace{0.5cm}
		\caption{Reconstruction of DM mass and $\mathcal{A}$ upon successful detection of DM in the future experiments, namely DARWIN only on right hand side and DARWIN combined with DarkSide-20k on the left hand side. The yellow and orange crosses marks the assumed true parameters that were used for generating the mock observations, \{$m_\chi = 30 \; \s{GeV}$, $\mathcal{A} = 2 \cdot 10^{-6} \, \s{GeV}^{-2}$\} and \{$m_\chi = 100 \; \s{GeV}$, $\mathcal{A} = 5 \cdot 10^{-6} \, \s{GeV}^{-2}$\}, respectively, while the contours correspond to the 68\% and 95\% credibility regions obtained under different assumptions regarding the local DM distribution. The dotted line marks current most stringent limits on $\mathcal{A}$ obtained by the XENON1T experiment.}
		\label{fig:dd_reconstruction_ana}
	\end{figure}

	The most significant advantages of the accurate phase-space modeling become apparent when considering DM with $m_\chi \gtrsim 100 \, \s{GeV}$ where the DM mass and coupling parameter become severely degenerate. The degeneracy can be to a certain extent mitigated by combining the results of experiments with different target nuclei, as can be seen by comparing the top and bottom panels of figure~\ref{fig:dd_reconstruction_si}, however, further notable improvements come from the use of refined astrophysical factors. From the bottom panel of the same figure, one can see that all of the equilibrium PSDFs perform significantly better in reconstructing the true model parameters than the SHM, which leads to more than 50\% larger uncertainties and a significant bias towards higher values of both parameters. Similar conclusions hold true also when considering DM coupled through the anapole moment -- the corresponding contour plots are shown in figure~\ref{fig:dd_reconstruction_ana} while the single-parameter confidence intervals are reported in table~\ref{tab:dd_reconstruction_ana}. The axisymmetric models allow for more accurate reconstruction of $m_\chi$ and the relevant cross-section due to two main reasons: firstly, the distinct knee in the astrophysical factors at intermediate $v_\s{min}$ (which is a consequence of anisotropic velocity distribution of DM, as described in section~\ref{sec:astrophysical_factors}) leads to a stronger dependence of the recoil spectrum on the DM mass and, secondly, they allow us to consistently take into account the correlation between $\rho(R_\odot)$ and DM's velocity distribution. Based on these arguments, as well as the results obtained for SI and anapole operators, it is very likely that the above-described trends in the reconstruction of DM parameters can be generalized to all possible DM-nucleon interaction types.
	
	\begin{table}
		\centering
		\renewcommand{\arraystretch}{1.3}
		\begin{tabular}{c||c|c||c|c}
			Model & $\log_{10} \left( \frac{m_\chi}{\s{GeV}} \right)$ & $\log_{10} \left( \frac{\mathcal{A}}{\s{GeV}^{-2}} \right)$ & $\log_{10} \left( \frac{m_\chi}{\s{GeV}} \right)$ & $\log_{10} \left( \frac{\mathcal{A}}{\s{GeV}^{-2}} \right)$ \\ \hline \hline
			True & 1.48 & -5.7 & 2 & -5.3 \\ \hline
			SHM (DARWIN only) & $1.47^{+0.08}_{-0.07}$ & $-5.68^{+0.04}_{-0.04}$ & $2.59^{+0.94}_{-0.63}$ & $-5.01^{+0.47}_{-0.29}$ \\
			NFW (DARWIN only) & $1.48^{+0.08}_{-0.06}$ & $-5.7^{+0.04}_{-0.04}$ & $2.55^{+0.97}_{-0.64}$ & $-5.03^{+0.48}_{-0.30}$ \\
			Burkert (DARWIN only) & $1.49^{+0.08}_{-0.06}$ & $-5.71^{+0.04}_{-0.04}$ & $2.56^{+0.96}_{-0.64}$ & $-5.04^{+0.48}_{-0.30}$ \\
			SHM (DARWIN \& DarkSide 20k) & $1.47^{+0.07}_{-0.06}$ & $-5.68^{+0.04}_{-0.04}$ & $2.02^{+0.22}_{-0.14}$ & $-5.28^{+0.1}_{-0.06}$ \\
			NFW (DARWIN \& DarkSide 20k) & $1.48^{+0.07}_{-0.06}$ & $-5.7^{+0.04}_{-0.04}$ & $2^{+0.14}_{-0.11}$ & $-5.3^{+0.06}_{-0.04}$ \\
			Burkert (DARWIN \& DarkSide 20k) & $1.49^{+0.07}_{-0.06}$ & $-5.71^{+0.04}_{-0.04}$ & $2.05^{+0.15}_{-0.12}$ & $-5.29^{+0.07}_{-0.04}$
		\end{tabular}
		\caption{True and reconstructed values of the DM mass and cross-section with the corresponding 68\% credibility intervals under different assumptions regarding the DM distribution function for the two benchmark scenarios with anapole interaction.}
		\label{tab:dd_reconstruction_ana}
	\end{table}
	
	The above tests were performed under the assumption that the interaction operator responsible for coupling between DM and nucleons is known. However, in practice, the task of accurately determining the properties of DM is even more difficult due to many possible ways in which DM can interact with nucleons. In order to address the effect of astrophysical factors on the capability of future experiments to distinguish among different scattering operators, we perform a test based on the Bayesian evidence ratios. Since in this work we are primarily interested in the implications that different DM distribution models have on the results of DD experiments, we consider only a subset of possible interaction types: the standard SI and SD interactions, coupling through $\mathcal{O}_{11}$ that arises in certain models with scalar mediators, anapole DM and millichared DM -- for more detailed discussion regarding these and other possible operators see, e.g.,~\cite{fitzpatrick_effective_2013,anand_model-independent_2013,dent_general_2015,gluscevic_identifying_2015,bishara_chiral_2017}. For a given model $\mathcal{M}$, corresponding to one of the possible interaction types between DM and baryons (for simplicity we neglect the possibility of more then one interaction type contributing to the signal), one can compute the Bayesian evidence by integrating the likelihood function~\eqref{eqn:dd_likelihood} over the entire parameter space $\Omega_{\mathcal{M}}$, weighted by the corresponding prior $p(\vec{\theta}|\mathcal{M})$:
	\begin{align}
	\label{eqn:evidence_def}
	\mathcal{Z}(\mathcal{D}|\mathcal{M}) = \int_{\Omega_{\mathcal{M}}} \mathrm{d}\vec{\theta} \; \mathcal{L}(\mathcal{D}|\vec{\theta},\mathcal{M}) \cdot p(\vec{\theta}|\mathcal{M}) \; .
	\end{align}
	In the above expression, $\vec{\theta}$ includes the unknown particle physics parameters, namely the DM mass and cross-section, as well as the unknown local DM distribution. Since we have no a priori knowledge regarding the DM mass and the coupling strength we adopt logarithmic non-informative priors for these parameters, i.e. $p(\log m_\chi|\mathcal{M}) = C_m$ and $p(\log \sigma_{\chi N}|\mathcal{M}) = C_\sigma$, where $C_m$ and $C_\sigma$ are the appropriate normalization constants. On the other hand, for astrophysical factors, the priors are determined by the posterior distribution of galactic parameters inferred from the kinematic data, as described in section~\ref{sec:mass_depomposition}. Despite these simplifications, the integral in equation~\eqref{eqn:evidence_def} is still not analytically tractable. Therefore, we approximate it using a Monte-Carlo integration method which allows us to rewrite it as a sum of the likelihood function evaluated in a large number of randomly picked points from the relevant parameter space $\Omega_{\mathcal{M}}$:
	\begin{align}
	\label{eqn:evidence_approx}
	\mathcal{Z}(\mathcal{D}|\mathcal{M}) \approx \frac{V_{\Omega_{\mathcal{M}}}}{N_\st{MC}} \sum_{i=1}^{N_\st{MC}} \mathcal{L}(\mathcal{D}|\vec{\Theta}_i,\mathcal{M}) \; .
	\end{align}
	In the above expression $V_{\Omega_{\mathcal{M}}} = C_m \cdot C_\sigma \cdot C_{f_\st{DM}}$ is the volume spanned by the parameter space $\Omega_{\mathcal{M}}$,~\footnote{In the expression for $V_{\Omega_{\mathcal{M}}}$ the factor $C_{f_\st{DM}}$ is related to the normalization of prior regarding the DM distribution function. However, $C_{f_\st{DM}}$ turns out to be irrelevant for the following discussion since it takes the same numerical value for all the interaction models and, hence, cancels out in quantities of interest for the model comparison.} $N_\st{MC}$ is the number of Monte-Carlo points used in the integration and $\Theta_i = \{\log m_{\chi}^{(i)}, \log \sigma_{\chi N}^{(i)}, f^{(i)}_\st{DM} \}$ is the parameter vector where $\log m_\chi^{(i)}$ and $\log \sigma_{\chi N}^{(i)}$ are uniformly sampled from the relevant intervals while $f_\st{DM}^{(i)}$ is the DM PSDF computed for a random sample from the aforementioned posterior distribution of galactic parameters. For a sufficiently large number of points (in our analysis we used $N_\st{MC} = 2 \cdot 10^5$ and explicitly checked that further increasing $N_\st{MC}$ does not change the results at the level of 1\% accuracy) this allowed us to obtain an accurate approximation of $\mathcal{Z}(\mathcal{D}|\mathcal{M})$. By computing the latter for all the considered interaction types, one can evaluate the probability that model $\mathcal{M}_i$ offers the optimal description of the data among the $N_\mathcal{M}$ considered models, assuming that initially all the models are equiprobable (i.e. $\s{Pr}(\mathcal{M}_i) = 1 / N_\mathcal{M}$)~\cite{wasserman_bayesian_2000}:
	\begin{align}
	\s{Pr}(\mathcal{M}_i|\mathcal{D}) = \frac{\mathcal{Z}(\mathcal{D}|\mathcal{M}_i)}{\sum_{j=1}^{N_\mathcal{M}} \mathcal{Z}(\mathcal{D}|\mathcal{M}_j)} \; .
	\end{align}
	As it was done in the case of parameter reconstruction, we included the intrinsic Poissonian noise by generating 512 stochastic realizations of the mock data using the same procedure as before. To obtain the final measure of the discrimination power for a given detector (or their combination), we studied the statistical properties of:
	\begin{align}
	\kappa \equiv \s{Pr}(\mathcal{M}_{\s{true}}|\mathcal{D}) \; ,
	\end{align}
	namely, the probability that a given realization of the mock data corresponds to the fiducial interaction type, which was used for generating the mock data.
	
\begin{figure}
		\centering
		\includegraphics[width=0.49\textwidth]{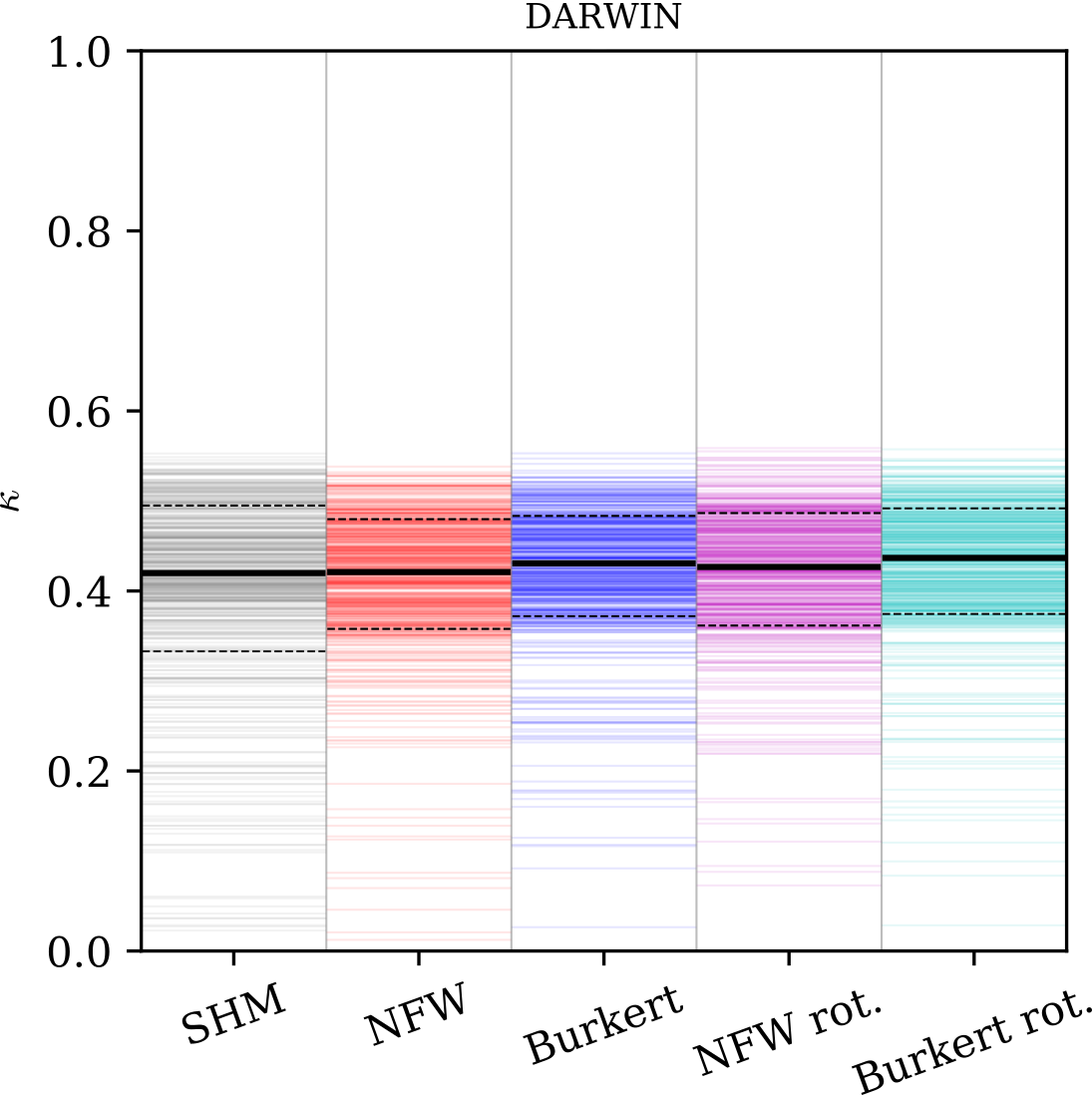}
		\includegraphics[width=0.47\textwidth]{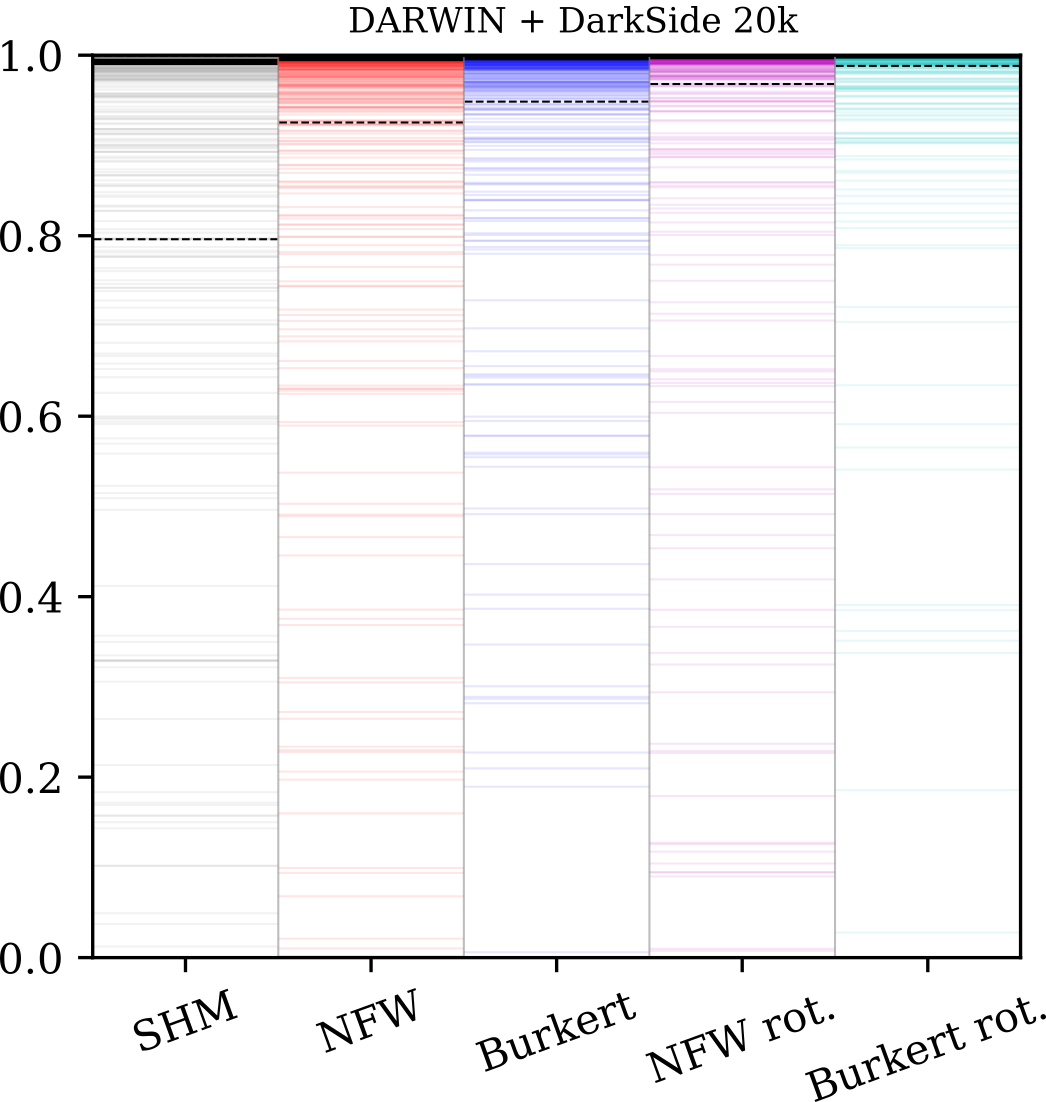}
		\vspace{0.5cm}
		\caption{The values of $\kappa$ computed for the mock datasets which were generated under the assumption of DM with $m_\chi = 100 \, \s{GeV}$ and $\sigma^\st{SI}_n = 5 \cdot 10^{-47} \, \s{cm}^2$. The black line marks the median value while the dashed lines mark the 16th and 84th percentiles.}
		\label{fig:dd_operator}
	\end{figure}

	The obtained values of $\kappa$ along with the corresponding 16th, 50th and 86th percentiles, obtained under the assumption of DM with $m_\chi = 100 \, \s{GeV}$ and $\sigma^\st{SI}_n = 5 \cdot 10^{-47} \, \s{cm}^2$, are displayed in figure~\ref{fig:dd_operator}. Similarly, as in the case of DM parameter reconstruction, the correct interaction type is difficult to identify by using DARWIN detector only, while the situation drastically improves when DARWIN and DarkSide-20k data are combined. For DARWIN experiment alone, all the astrophysical models perform equally well on average, however, the SHM shows roughly 10\% larger spread in $\kappa$ around the central value than the rest of the considered models. This implies that the identification of the scattering operator under the assumption of SHM is more susceptible to the intrinsic fluctuations in the observational data. Upon combining the signals of DARWIN and DarkSide-20k detectors, a much larger discrepancy between SHM and equilibrium models becomes evident. The median value of $\kappa$ for the SHM is around 0.992 while for equilibrium models it reaches 0.999 or higher. Even more importantly, one can observe large differences at the lower end of the distribution of $\kappa$, where 84\% of stochastic datasets lead to $\kappa > 0.79$ in the case of the SHM while for the equilibrium models 84\% of stochastic datasets lead to $\kappa > 0.93$ or better. In other words, under the assumption of the SHM correct categorization of scattering operator is much more sensitive to statistical fluctuations in the event count, even for relatively large values of cross-section that are only slightly below the current limits, while the kinematically constrained equilibrium models lead to more accurate categorization, regardless of the assumed DM density profile and halo's rotational properties. Similarly as in the case of parameter reconstruction, these differences can be mainly attributed to more distinct features in the differential event rate, which arise due to the pronounced knee in the astrophysical factors of equilibrium models, and the inclusion of correlations between the local DM density and its velocity distribution. For lower DM masses we find that the identification of the scattering operator becomes more difficult (as was already shown in, e.g.,~\cite{edwards_dark_2018}), and hence the astrophysical factors only play a minor role. As a side note, it is perhaps unexpected to see that the NFW model, which was used for generating the mock data, does not lead to the most reliable identification of the scattering operator. However, this is simply a consequence of the fact that the mock data provide an even worse match to other operators if alternative equilibrium distribution functions are assumed. On the other hand, we explicitly checked that by fixing the scattering operator to SI and comparing the evidence of different phase-space models, the largest probability is given to the NFW one, as expected.
	
	\subsubsection{Annual modulation}
	\label{sec:direct_detection_modulation}
	
	As it was already mentioned above, annual modulation of the nuclear recoil rate represents a valuable handle for corroborating the DM origin of DD events~\cite{spergel_motion_1988,freese_signal_1988,freese_colloquium_2013,froborg_annual_2020}. The modulation arises since the mapping of DM velocity into the LAB frame, given by equation~\eqref{eqn:v_lab}, varies throughout the year due to time dependent orientation of $\vec{v}_\s{circ}(t)$. The circular motion of the Earth, therefore, induces an approximately sinusoidal yearly variations of the expected recoil rate with its extremes roughly corresponding to $\vec{v}_\s{circ}(t)$ being aligned or counter-aligned with the $\vec{v}_\st{LSR} + \vec{v}_\odot$, which occurs around 1st of June and 1st of December. While the shape and phase of annual modulation are primarily determined by the Earth's orbit around the Sun, its amplitude can be significantly affected by the assumptions regarding the DM's velocity distribution.
	
\begin{figure}
		\centering
		\includegraphics[width=0.49\textwidth]{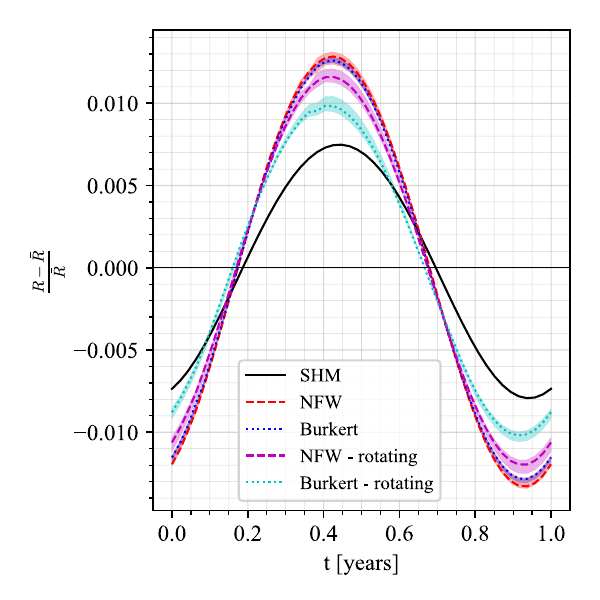}
		\includegraphics[width=0.49\textwidth]{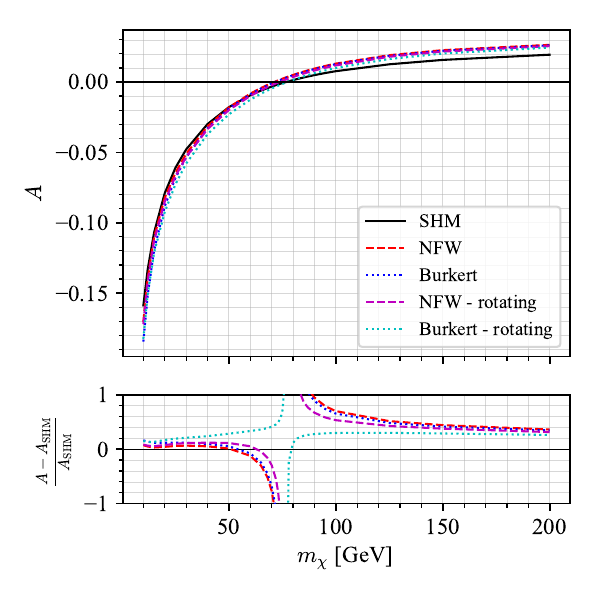}
		\caption{\textit{Left-hand side}: The expected nuclear recoil rate in Xenon-based experiments, normalized to the yearly average, as a function of time for the considered benchmark halo models with the corresponding 68\% h.p.d. regions. \textit{Right-hand side}: The median expected amplitude of annual modulation in Xenon-based experiments as a function of DM mass.}
		\label{fig:modulation}
	\end{figure}

	In order to quantitatively explore these differences we re-evaluate the astrophysical factors for 1000 randomly picked samples of the galactic parameters from the MCMC scan (discussed in section~\ref{sec:mass_depomposition}) at different times $t$, which allowed us to obtain the posterior distribution of the expected recoil rate throughout the year. In the left-hand side plot of figure~\ref{fig:modulation} we show the yearly evolution of the expected recoil rates, normalized by the yearly average, in Xenon-based experiment for our benchmark PSDF models, assuming $m_\chi = 100 \; \s{GeV}$ and SI DM-nucleus interactions. As can be seen from the plot, the axisymmetric models predict a significantly (between 30\% and 50\%) larger amplitude of the annual modulation which is mainly a consequence of the velocity anisotropy of DM particles that is induced by the presence of the baryonic disc -- as it was mentioned before, in axisymmetric models the disc reduces the velocity dispersion of DM along the azimuthal direction, while the velocity dispersion in the meridional plane is increased~\cite{petac_two-integral_2019}. Consequently, the predicted recoil rates are more sensitive to the variation of Earth's velocity along the azimuthal direction, which roughly coincides with the $\vec{v}_\st{LSR} + \vec{v}_\odot$, leading to increased yearly modulation of the signal. Significant differences can also arise due to the assumptions regarding the halo's rotation. The non-rotating equilibrium models predict the highest modulation amplitude, with significant overlap of their corresponding 68\% h.p.d. bands, while somewhat lower yearly variations were found for the co-rotating halos.~\footnote{The opposite is true for counter-rotating halos, i.e., they lead to even higher modulation amplitudes than the non-rotating models. However, we chose to exclude this option from our analysis since counter-rotating halos are very rarely found in numerical simulations of structure formation and we currently have no evidence in favor of such exceptional merger history in the case of the Milky Way's DM halo.} In the latter case, the amplitude of modulation clearly depends on the rotational velocity, which is the main reason for the difference between the two considered DM density profiles -- by imposing $r_a = r_s$ and $\lambda = 0.04$, as explained in section~\ref{sec:DM_assumptions}, the Burkert density profile leads to larger $\bar{v}_\phi(R_\odot)$, and consequently lower amplitude of the annual modulation, in comparison with the NFW case. Apart from the modulation amplitude, some other small differences among the predictions of various PSDFs can be observed. The equilibrium models tend to reach the maximum/minimum recoil rate a little earlier than the SHM, but also the sinusoidal shape of the modulation signal can get slightly distorted around the peak. Both of these effects are again consequences of the DM's orbital anisotropy and halo's rotational properties, which makes them most clearly visible in the case of the rotating Burkert halo. However, it has to be noted that these features depend on the assumed DM mass and are noticeable only around $m_\chi \sim 75 \; \s{GeV}$, where, on the other hand, the amplitude of modulation is negligible, which makes them extremely difficult to observe in practice. What is experimentally more important, is the dependence of modulation amplitude on the DM mass:
	\begin{align}
	A(m_\chi) = \frac{R(t_1; m_\chi) - R(t_2; m_\chi)}{R(t_1; m_\chi) + R(t_2; m_\chi)} \; ,
	\end{align}
	where $R(t; m_\chi)$ is the total expected recoil rate at time $t$ for DM with mass $m_\chi$, while $t_1$ and $t_2$ correspond to the 1st of June and 1st of December. We plot the modulation amplitude as a function of DM mass in the right-hand side of figure~\ref{fig:modulation}, from where it can be observed that the amplitude vanishes around $m_\chi \approx 75 \; \s{GeV}$, while it grows with increasing and decreasing DM mass, but has the opposite sign. The low mass regime also exhibits much larger modulation amplitudes, which can exceed 10\% of the total recoil rate, while for large DM masses they remain below 5\%. In the bottom panel of the same plot, where we show the ratio of expected modulation amplitude in equilibrium models and the SHM, it can be seen that the relative differences are at the order of $10\%$ at low DM masses and more than 30\% at large DM masses. Even larger ratios of expected amplitudes can arise around $m_\chi \sim 75 \; \s{GeV}$, however, due to the vanishing amplitude these distinctions are again most likely unobservable. From the bottom panel, one can additionally notice that for all the considered equilibrium models the phase inversion occurs at slightly lower DM masses with respect to the SHM, except for the rotating Burkert halo, for which it does at slightly larger DM mass.
	
	Our results clearly show the need of using axisymmetric models for accurately predicting the annual variation of the recoil rate. While the predictions of non-rotating equilibrium models are fairly independent of the assumed DM density profile, the halo's rotational properties can have a significant impact on the expected amplitude of signal modulation. Since $\bar{v}_\phi(R,z)$ is observationally unconstrained, it can be understood as a source of irreducible uncertainty in the astrophysical factors and should be marginalized over. While this falls beyond the scope of our current analysis, we non-the-less provide tabulated values of $g(v_\s{min})$ and $h(v_\s{min})$ at different times throughout the year for the considered axisymmetric models, which can be used to obtain representative estimates of the expected annual modulation for an arbitrary DD experiment and DM candidate.
	
	\subsubsection{Directional detection}
	
	Finally, we turn our attention to the directional distribution of DM-induced nuclear recoils. This will be of the prime interest for the future DD experiments that will be capable of reconstructing the orientation of scatterings in addition to the over-all rate and its energy spectrum -- for a review of various suggested experimental designs see, e.g.,~\cite{mayet_review_2016,couturier_dark_2017,vahsen_cygnus_2020}. One of the key reasons for the development of such detectors is their ability to further discriminate between genuine DM signals from otherwise irreducible backgrounds. Typical examples of such backgrounds are solar, atmospheric, and supernovae neutrinos, which are expected to induce nuclear recoils with similar energy spectra as DM. Currently, existing detectors that are capable of reconstructing the direction of recoils have very poor angular resolution and/or inferior sensitivity (note that none of the DD experiments has yet reached the sensitivity needed to detect any of the aforementioned neutrino sources).
	Since the directional distribution of events strongly depends on the local velocity distribution of DM, we take a closer look at the predictions of axisymmetric models and compare them with the ones obtained from the SHM.
	
	In the context of directional detection experiments the key quantity to consider is the double differential recoil rate~\cite{gondolo_recoil_2002}:
	\begin{align}
	\frac{\s{d}^2R}{\s{d}E_r \; \s{d}\Omega}(E_r, \hat{q}) = \frac{1}{m_A m_\chi} \cdot \int \s{d}^3v \; f(\vec{x}, \vec{v}) \cdot v \cdot \frac{\s{d}^2 \sigma}{\s{d} E_r \; \s{d}\Omega} \label{eqn:rate_double_dif} \; ,
	\end{align}
	where $\hat{q}$ is the orientation vector, $\frac{\s{d}^2 \sigma}{\s{d} E_r \; \s{d}\Omega}$ is the double differential DM-nucleus cross-section while other quantities are the same as in equation~\ref{eqn:rate_dif}. Up to the second order in momentum transfer and relative DM-nucleus velocity, all the information regarding the angular distribution of the nuclear recoils can be again fully encoded in two characteristic functions, see, e.g.,~\cite{catena_dark_2015,kavanagh_new_2015,mayet_review_2016}. For brevity, we demonstrate the effect of axisymmetric modeling on directional distribution of events by studying the astrophysical factor corresponding to the most common $\s{d}\sigma/\s{d}E_r \propto v^{-2}$ case, which we define as:
	\begin{align}
	i(v_\s{min}, \hat{q}) & \equiv \frac{1}{g(v_\s{min})} \int \s{d}^3v \; f(\vec{x}, \vec{v}) \cdot \delta \left( \vec{v} \cdot \hat{q} - v_\s{min} \right) \; .
	\end{align}
	Analogously to the annual variation of $g(v_\s{min})$, also $i(v_\s{min}, \hat{q})$ is expected to vary throughout the year due to the time dependence in mapping between the DM halo's and LAB's rest frames.
	
	\begin{figure}[h]
		\centering
		\includegraphics[width=\textwidth]{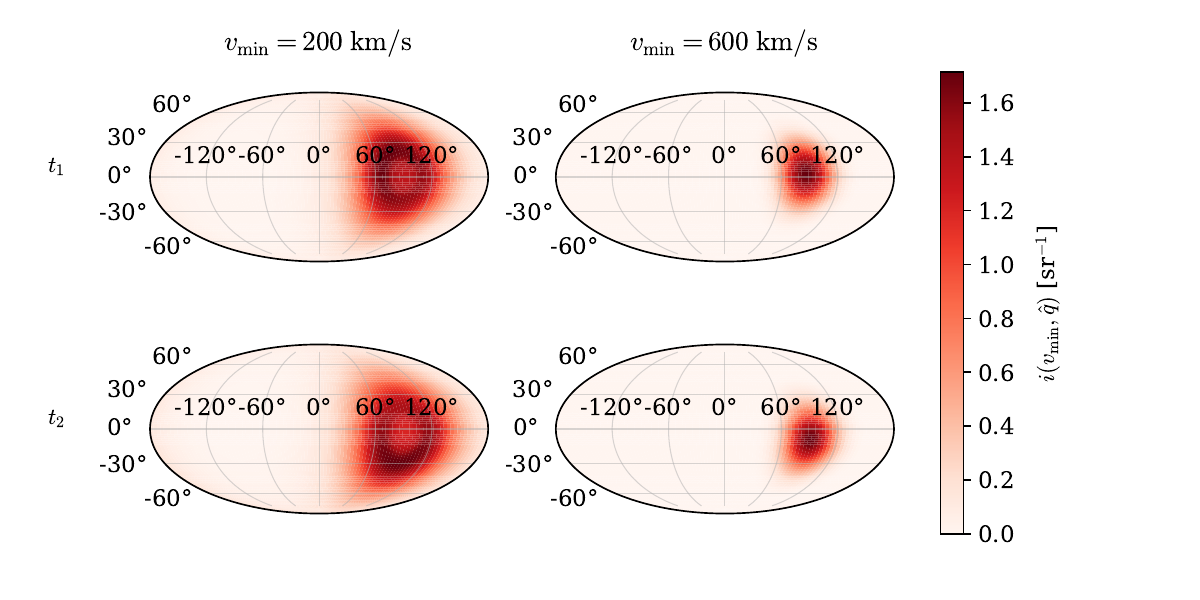}
		\caption{Directional distribution of recoil events, normalized to the total rate. The left column corresponds to recoils with $v_\s{min} = 200 \; \s{km/s}$ and the right column to $v_\s{min} = 600 \; \s{km/s}$. The top row is the directional distribution expected around 1st of June and bottom around 1st of December.}
		\label{fig:direction_nfw}
	\end{figure}
	
	In figure~\ref{fig:direction_nfw} we show the directional distribution of nuclear recoils obtained for the non-rotating axisymmetric model with NFW density profile at the yearly minimum ($t_1$) and maximum ($t_2$). As can be seen from the plots, which are a Mollweide projection of the sky sphere with origin pointing towards the galactic center, $i(v_\s{min}, \hat{q})$ exhibits a characteristic monopole pattern since the majority of events are expected from the direction of Earth's relative movement with respect to the galactic rest frame, roughly coinciding with the position of the Cygnus constellation. The position and shape of the peak slightly varies with the time of the year, however, the differences are relatively small and very good angular resolution would be required to detect them. On the other hand, the directional distribution of nuclear recoils strongly depends on $v_\s{min}$. For values well below the local escape velocity the highest recoil rates are expected in a circular band around the direction of the Cygnus constellation, forming a ring-like feature that was previously noted in, e.g.,~\cite{kavanagh_new_2015,mayet_review_2016}, while for larger values the ring disappears and the maximum becomes aligned with the Earth's motion relative to the galactic rest frame. We find the same behavior as described above for all the considered halo models, however, appreciable differences between them can still arise.
	
\begin{figure}[h]
		\centering
		\includegraphics[width=\textwidth]{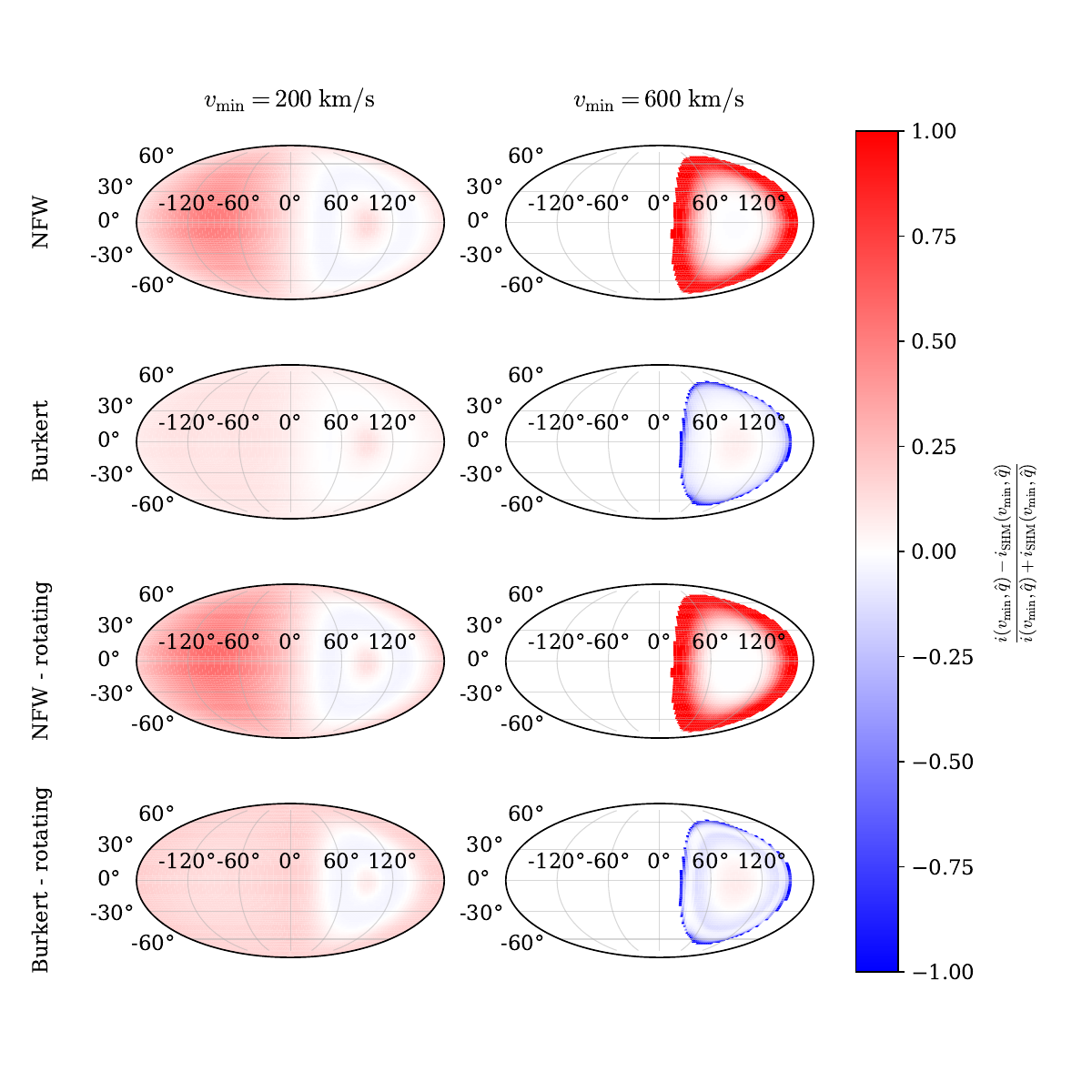}
		\caption{Yearly-averaged difference of normalized directional distribution of nuclear recoils between the equilibrium models and the SHM. The left column corresponds to recoils with $v_\s{min} = 200 \; \s{km/s}$ and the right column to $v_\s{min} = 600 \; \s{km/s}$.}
		\label{fig:direction_diff}
	\end{figure}

	In order to emphasize the impact of various modeling assumptions we plot in figure~\ref{fig:direction_diff} the yearly-averaged difference between each of our four axisymmetric models and the SHM, normalized by the sum of $i(v_\s{min}, \hat{q})$ of the two models. From there one can see that the models based on NFW density profile lead to a more uniform distribution of events; for $v_\s{min} \lesssim v_\s{esc}$ the event rate in the ring-like band tends to be slightly smaller while somewhat larger numbers of recoils are expected from the direction of the Cygnus constellation as well as from the other side of the sky hemisphere. While the differences are at the order of 10\% for the non-rotating halo, they can become up to 25\% for the rotating halo since the number of events coming from the opposite direction of Earth's movement is significantly increased. At larger $v_\s{min}$ the NFW model again leads to a more uniform distribution of events, but in this case the peak is nearly unaffected while the number of recoils around it is increased, which also leads to a larger region of the sky from which events can originate. On the contrary, for Burkert models the characteristic peak at large $v_\s{min}$ becomes even sharper while the patch of sky from which the events originate shrinks. At small $v_\s{min}$ the trends of Burkert models are more similar to the ones of NFW density profile -- the ring feature is nearly unchanged, however, at its center and outside of it a slightly higher recoil rate is expected. On the celestial hemisphere opposite to the Earth's movement direction the change in expected number of events most strongly depends on the rotational properties of the DM halo. For non-rotating halos the recoil rate is decreased by roughly 10\% while for the rotating halo a few percent increase can be observed. While further differences can arise by, e.g., considering a broader set of scattering operators, focusing on the exact pattern of yearly variations, more carefully studying the $v_\s{min}$ (i.e. recoil energy) dependence or marginalizing over all possible $\bar{v}_\phi(R,z)$, we in this work restrain ourselves from such a detailed analysis.
	
	\section{Implications for indirect detection}
	\label{sec:indirect_detection}
	
	If DM particles can annihilate in SM states, which is generically true for thermal relic candidates, the associated emissions could be detected through various messengers, ranging from $\gamma$-rays, neutrinos to cosmic rays. Currently the dominant limits on most annihilation channels come from $\gamma$-ray observations of the galactic center~\cite{collaboration_search_2018,chang_search_2018,fermi-lat_collaboration_search_2019,abazajian_strong_2020} and dwarf satellite galaxies~\cite{collaboration_dark_2014,ackermann_searching_2015,archambault_dark_2017,boddy_model-independent_2018,hoof_global_2018,the_hawc_collaboration_search_2020,rico_gamma-ray_2020,alvarez_dark_2020}, however, also interesting bounds can also be obtained in combinations with the measurements of cosmic ray fluxes~\cite{boudaud_robust_2019,belotsky_cosmic_2020}. The results of this work are the most important in the context of possible DM annihilation signals originating from the galactic center (GC), which crucially depend on the distribution of DM within the inner parts of our galaxy. The expected flux due to pair annihilation of DM into SM particles, in the following denoted as $\psi$, for a given DM halo with PSDF $f(\vec{x}, \vec{v})$ integrated over the angular acceptance $\Delta \Omega$, is given by:
	\begin{align}
	\label{eqn:gen_annihilation_flux}
	\frac{d \Phi_{\psi}}{dE_{\psi}} = \frac{1}{8 \pi} \frac{\sv_0}{m^2_{\chi}} \frac{dN_{\psi}}{dE_{\psi}} \, \int_{\Delta \Omega} d \Omega \int_{\textrm{l.o.s.}} d \ell \int d \vec{v}_1 f (\vec{x}, \vec{v}_1) \int d \vec{v}_2 \;  f(\vec{x}, \vec{v}_2) \, S( | \vec{v}_{\textrm{rel}}|) \,,
	\end{align}
	where the DM particle $\chi$ is assumed to be its own antiparticle (otherwise an extra factor of 1/2 is needed), $m_{\chi}$ is its mass and $dN_{\psi}/dE_{\psi}$ the energy spectrum of the produced $\psi$ particles per annihilation. The above formula is applicable to the general case in which the pair annihilation cross-section $\sv$ has a non-trivial dependence on the modulus of the relative velocity $v_\s{rel} = |\vec{v}_1 - \vec{v}_2|$, with $\vec{v}_1$ and $\vec{v}_2$ being the velocities of two annihilating particles -- therefore, $\sv$ is factorized into the velocity independent term $\sv_0$ times a dimensionless factor fully comprising its dependence on relative velocity, $\sv = \sv_0 \cdot S(v_\s{rel})$.
	
	By isolating the astrophysical contribution in equation~\eqref{eqn:gen_annihilation_flux}, one can define:
	\begin{align}
	\label{eqn:j_factor}
	J \equiv & \int_{\Delta \Omega} d \Omega \int_{\textrm{l.o.s.}} d \ell \int d \vec{v}_1 f_\st{DM} (\vec{x}, \vec{v}_1) \int d \vec{v}_2 \;  f_\st{DM}(\vec{x}, \vec{v}_2) \, S(v_\s{rel}) \nonumber\\
	= & \int_{\Delta\Omega} d\Omega \int_{\rm l.o.s.} d\ell \, \rho^2(\vec{x}) \ \langle S (v_\s{rel}) \rangle (\vec{x}) \, .
	\end{align}
	This definition is in analogy to what is usually denoted in the literature as ``$J$-factor'', which is typically limited on the standard lore of s-wave annihilations, where $\sv$ is velocity independent.
	In the latter case, the velocity boost factor can be omitted, i.e. $\langle S (v_\s{rel}) \rangle = 1$.
	The lack of tangible signals from the simplest WIMP scenarios motivates the exploration of more general models in which non-trivial velocity dependence in the thermally averaged cross-section can arise. For example, there exist theoretical proposals in which s-wave annihilations are forbidden or severely suppressed, and hence p-wave processes become relevant~\cite{hagelin_perhaps_1984,kim_minimal_2007,pospelov_secluded_2008,lee_singlet_2008}, leading to $S(v_\s{rel}) \propto v_\s{rel}^2$. Alternatively, non-perturbative effects due to long-range interactions in the non-relativistic limit, commonly known as Sommerfeld enhancement, can introduce an additional velocity dependence, which can be in certain limiting cases well approximated by inverse powers of $v_{\textrm{rel}}$~\cite{iengo_sommerfeld_2009,dent_thermal_2010,slatyer_sommerfeld_2010,tulin_beyond_2013}. However, it is important to note that in the modeling of GC emissions associated with DM annihilation one of the main sources of uncertainty is the DM density distribution in the very center of our galaxy. Therefore, we in the following first turn our attention to the values of $J$-factor under the assumption of $S(v_\s{rel}) = 1$ and subsequently consider a range of possible velocity dependences that can arise in the annihilations processes. We again obtain central values and h.p.d. regions from the sampling of the galactic mass model described in section~\ref{sec:mass_depomposition}, but this time our results need to be taken with a degree of caution; since the innermost data used in our analysis lies at $R \sim 2 \; \s{kpc}$, the derived quantities are extrapolations of our model and are primarily driven by our assumptions regarding the DM density profiles. Furthermore, around the GC, the dynamics are dominated by the bulge, which exhibits triaxial morphology and, therefore, can not be fully captured by the axisymmetric distribution functions used in this work. On the other hand, due to the lack of observational constraints and the complex structure of the region, we believe that our results still provide valuable estimates for the astrophysical factor associated with the GC.
	
	\subsection{Astrophysical factors for s-wave annihilations}
	
	For annihilations with velocity independent thermally averaged cross-section the $J$-factors are fully determined by the DM density profile. However, the distribution of DM near the GC is subjected to large uncertainties since it constitutes only a small fraction of the dynamical mass while, at the same time, performing accurate observations is extremely challenging due to the complexity of the region. Furthermore, the central DM density slope can significantly deviate from the standard parametric forms due to, e.g., the central super-massive black hole~\cite{gondolo_dark_1999,ullio_dark-matter_2001,merritt_dark_2002,bertone_annihilation_2002,gnedin_dark_2004,vasiliev_dark_2007,lacroix_intermediate-mass_2018} or baryonic effects~\cite{chan_impact_2015,benitez-llambay_baryon-induced_2019,artale_dark_2019}. On the other hand, analyses of GC emissions often mask out the inner $2^\circ$ around the GC, beyond which our benchmark models should provide more reliable estimates on the DM density, however, the inferred astrophysical factors still strongly depend on the assumption regarding central slope of the DM density profile. For the convenience of future analysis we express our results in terms of the differential $J$-factor with respect to the aperture angle $\alpha$:
	\begin{align}
	\label{eqn:J_diff}
	\frac{\s{d}J}{\s{d}\alpha} = 2 \pi \sin \alpha \int_0^{\infty} d\ell \, \rho^2\left(r (l,\alpha) \right) \quad \s{where} \quad r(l,\alpha) = \sqrt{l^2 + R_\odot^2 - 2 l R_\odot \cos \alpha} \; .
	\end{align}
	
\begin{figure}[h]
		\centering
		\includegraphics[width=5in]{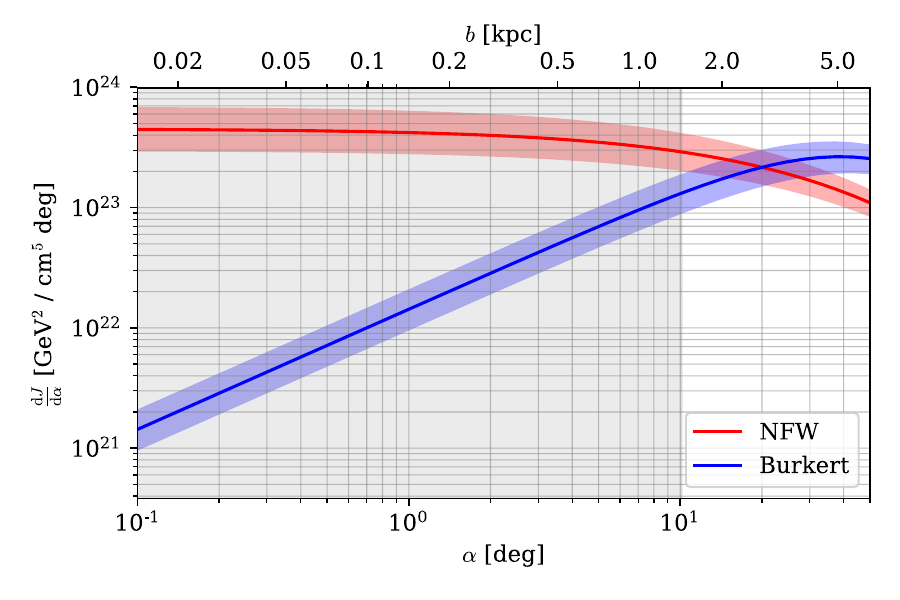}
		\caption{Differential $J$-factors and their corresponding 68\% h.p.d. band as a function of angle between GC and the line-of-sight (or the corresponding impact parameter specified in the top axis).}
		\label{fig:dj}
	\end{figure}

	In figure~\ref{fig:dj} we show the central values of the above quantity for our benchmark (NFW and Burkert) density profiles together with the 68\% h.p.d. regions as derived from the sampling of galactic mass models. As can be seen from the plots, there is a significant difference between the predictions of cuspy and cored DM profiles. For $\alpha \lesssim 10^\circ$ the NFW model leads to significantly larger and nearly constant values of differential $J$-factor, while for the Burkert case it decreases inversely proportional to $\alpha$, which leads to values that can be more than an order of magnitude below the NFW prediction for $\alpha \lesssim 3^\circ$. On the other hand, at $\alpha \approx 20^\circ$ the differential $J$-factors of the two models coincide, and the annihilation flux can be predicted independently of the assumed density profile. At even larger values of $\alpha$ the Burkert, profile leads to significantly higher $\frac{\s{d}J}{\s{d}\alpha}$, again making the two models inconsistent by a large margin.
	These discrepancies in the predicted differential $J$-factors shall be interpreted as unaccounted systematic error that arises from the assumptions regarding the parametric form of the DM density profile. This could be mitigated by using tighter kinematic constraints for the inner part of the galaxy, but in such analysis also $\rho$ with a variable central slope should be used. While such endeavor is beyond the scope of this work, we conclude that our results for the Burkert profile could be taken as a fairly robust lower bound on the differential $J$-factor for $\alpha \lesssim 10^\circ$. On the contrary, the NFW profile does not necessarily provide an upper limit in this range (neither a lower limit at $\alpha \gtrsim 20^\circ$) since even steeper central cusp could be present in the our galaxy. Finally, it is also worth noting that the contribution to the $J$-factor originating from $\alpha \sim 20^\circ$ minimally depends on the assumed DM density profile and could be used to constrain the DM's annihilation cross-section without relying on a particular parametric form of the DM density profile.
	
	\subsection{Velocity boost factors}
	
	In case of non-trivial velocity dependence of the annihilation cross-section one needs to additionally take into account the velocity boost factor $\langle S(v_\s{rel}) \rangle$. The latter is obtained by averaging the cross-section's velocity dependence over the relative velocity distribution dictated by the PSDFs of the annihilating particles. As it was already mentioned above, in the context of indirect detection the most commonly considered velocity dependences are due to p-wave annihilations with $S(v_\s{rel}) \propto v_\s{rel}^2$ or due to Sommerfeld enhancement, which can be under the assumption of the Yukawa coupling approximated as $S(v_\s{rel}) \propto v_\s{rel}^{-1}$ in the Coulomb regime (i.e. for vanishing mediator mass) and as $S(v_\s{rel}) \propto v_\s{rel}^{-2}$ in the resonant regime (occurring for particular values of the associated particle physics parameters). This motivates us to consider four different power-law scalings, namely $S(v_\s{rel}) = v_\s{rel}^{\xi}$ for $\xi \in [2,1,-1,-2]$, corresponding to all possible combinations of the aforementioned velocity dependences.
	
\begin{figure}[h]
		\centering
		\includegraphics[width=\linewidth]{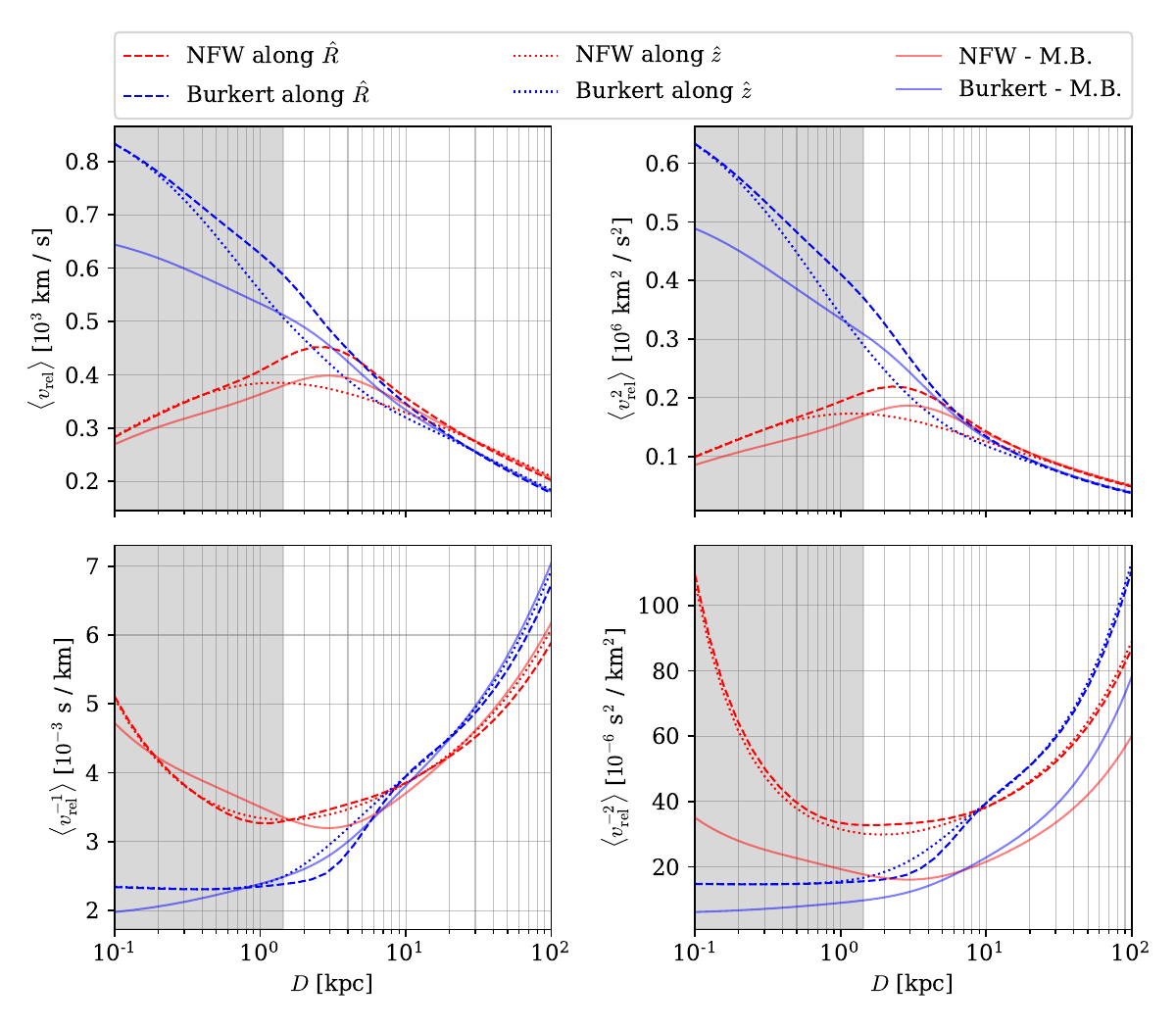}
		\caption{Moments of the relative velocity distribution as a function of the galactocentric distance. For axisymmetric models the distinct values along the $\hat{R}$ and $\hat{z}$ axes are plotted.}
		\label{fig:vrel_moments}
	\end{figure}

	In figure~\ref{fig:vrel_moments} we show the corresponding moments of the relative velocity distribution evaluated at different galactocentric distances for the PSDFs obtained from the central values of the galactic parameters, reported in section~\ref{sec:mass_decomposition_results}. We present the results for both models, assuming either a NFW or Burkert density profile, as well as the corresponding predictions of the SHM with a velocity dispersion profile derived from the spherical Jean's analysis (for details see, e.g.,~\cite{binney_galactic_2008}):
	\begin{align}
	\sigma^2(r) = \frac{1}{\rho(r)} \int_r^{\infty} \s{d}r' \rho(r') \frac{\s{d} \Psi^\st{SPH}_\s{tot}}{\s{d}r}(r) \, .
	\end{align}
	Since the gravitational potential that was used to fit the kinematic data is axisymmetric, the corresponding spherical gravitational potential $\Psi^\st{SPH}_\s{tot}(r)$ has to be determined for the SHM. This can be achieved by demanding that the associated rotation curve remains unchanged, which can be formally expressed as:
	\begin{align}
	\frac{\s{d} \Psi^\st{SPH}_\s{tot}}{\s{d}r}(r) \equiv \left. \frac{\s{d} \Psi_\s{tot}}{\s{d}R}(R,z) \right|_{\substack{R = r \\ z=0}}
	\end{align}
	On the other hand, for axisymmetric models we derive the distinct predictions of $\langle v_\s{rel}^\xi \rangle$ along $\hat{R}$ and $\hat{z}$ axes.~\footnote{In case of Eddington's inversion the relative velocity moments would be similar to those of axisymmetric model, however, with small differences depending on the prescription used for transforming the axisymmetric gravitational potential into a spherical one. For example, if the rotation curve was to be preserved, its predictions would be very close to those of axisymmetric model along $\hat{R}$ axis, while if it the enclosed mass within a spherical radius $r$ was preserved, they would lie closer to the prediction for $\hat{z}$ axis.} From the plots one can see that the spherical Jean's approximation performs relatively well for large galactocentric distances, $D$, relative to the disc scale length (i.e. at $D \gg a_\s{disc}$), except in the case of $\langle v_\s{rel}^{-2} \rangle$ which leads to significantly lower values of $\langle S(v_\s{rel}) \rangle$ under the assumption of Maxwell-Boltzmann velocity distribution (this remains true for all $D$). At large $D$ also the difference between equilibrium models, assuming either NFW or Burkert DM density profile, is relatively small. The NFW model typically leads to a relative velocity distribution of DM particles that is shifted to slightly higher velocities, which results in increased positive moments and decreased negative moments of $v_\s{rel}$. However, at galactocentric distances comparable to the scale length of the baryonic disc, larger differences between the considered models arise. First of all, in the axisymmetric approach, one can see a significant impact of the baryonic disc on the dynamics, which reflects in different values of the relative velocity moments along $\hat{R}$ and $\hat{z}$ axes, where the former is associated with higher values of positive moments and lower values of negative moments. On the other hand, the SHM model, which is built on the assumption of spherical symmetry, typically leads to values of $\langle v_\s{rel}^\xi \rangle$ that fall between the expectations for the corresponding axisymmetric model along different axes. At galactocentric distances smaller than the innermost kinematic measurement used in this work (gray-shaded region), the galactic bulge begins to dominate the dynamics, which results in a decreasing difference between the relative velocity moments along different axes of the axisymmetric models. However, it should be kept in mind that this is a consequence of our assumption of a spherical bulge and we expect that it could be appreciably modified in more realistic setups. For the model based on the NFW density profile also the difference with respect to the SHM model decreases (except for $\langle v_\s{rel}^{-2} \rangle$), while the opposite is true for the Burkert profile. However, the most significant contrasts arise due to the assumptions regarding the central density slope, since the cuspy NFW profile leads to appreciably lower relative velocities of DM particles at small galactocentric distances, where the bulk of the signal is expected to originate, in comparison with the cored profile. This results in velocity boost factors that are a few times larger in the case of the Burkert halo for $\langle v_\s{rel}^\xi \rangle$ with $\xi > 0$ and are a few times lower for $\xi < 0$, in comparison with the predictions for the NFW halo. Upon convolving the positive relative velocity moments with the corresponding $\rho^2(r)$ along the line of sight, a mild cancellation between increased DM density and lower values of $\langle S(v_\s{rel}) \rangle$ might occur for NFW  case. In contrast, for negative velocity moments, the difference between cuspy and cored profile is expected to become even larger. These results further emphasize the importance of correctly determining the central DM density slope for accurately predicting the annihilation flux originating from the GC, even in the case of velocity-dependent annihilations.
	
	\section{Summary and conclusions}
	\label{sec:conclusions}
	
	Modern astronomical observations and ever-increasing sensitivity of DM searches compel us to improve the modeling of the DM distribution within our galaxy. While in the past crude but computationally friendly approximations were often used, we provide refined halo models which most noticeably differ from previous works by allowing for a self-consistent axisymmetric description of the galactic halo under the assumption of dynamical equilibrium. Since the dynamics within the inner $\sim 10 \; \s{kpc}$ of the Milky Way are dominated by the flattened baryonic disc, and it is generically expected that DM halos exhibit some degree of net rotation around the central axis, our approach provides important refinements of the predictions for astrophysical factors which crucially enter the interpretation of DM searches. Below we summarize our key results regarding the structure of our galaxy and its implications for direct and indirect DM searches.
	
	\paragraph*{Galactic mass decomposition:}
	
	In the first part of this work, we introduced the method used for computing the axisymmetric PSDF of equilibrium systems and discussed the available kinematic measurements that can be utilized to constrain such models within the Milky Way. Subsequently, we used a set of latest observations to sample two benchmark models, assuming either a cuspy NFW or a cored Burkert density profile, through the Bayesian MCMC approach. By using fairly general parametric functions for describing baryonic components of our galaxy (namely the bulge and the disc) and generous priors on the related parameters, we obtained robust constraints on the considered galactic mass models which are in good overall agreement with previous studies. Our results show that both, NFW and Burkert, halo models provide a good fit to the kinematic tracers in the range of radial distances where the data is most constraining, $5 \; \s{kpc} \lesssim R \lesssim 20 \; \s{kpc}$, even though there is a slight preference in favor of the NFW model that is primarily driven by the observations in the outskirts of the galaxy. Not surprisingly, the obtained DM density in the aforementioned range is similar in both models with significant overlap of the corresponding 68\% credibility bands. For the local DM density we found $\rho_\st{NFW}(R_\odot) = 0.941^{+0.053}_{-0.057} \cdot 10^{-2} \; \s{M}_\odot / \s{pc}^3$ and $\rho_\st{BUR}(R_\odot) = 1.00^{+0.054}_{-0.057} \cdot 10^{-2} \; \s{M}_\odot / \s{pc}^3$ under the assumption of NFW and Burkert profile, respectively. Also, the posterior distributions of the parameters related to the baryonic components are fairly consistent apart from the mass and scale length of the baryonic disc. The latter tends to be slightly less massive and more extended in the case of the NFW density profile, which can have intricate consequences for the interpretation of direct detection experiments.
	On the other hand, the deployed kinematic observations, which are limited to galactocentric distances $D \gtrsim 2 \; \s{kpc}$, are not sufficient to accurately determine the DM distribution within the central few $\s{kpc}$ of our galaxy, which is of the prime interest for indirect searches.
	
	\paragraph*{Direct detection:}
	
	After obtaining the posterior distributions of parameters for the two benchmark galactic models, we turned our attention to their implications for the DM searches. First we examined their impact on direct detection experiments which are particularly sensitive to the local DM density and its velocity distribution through the astrophysical factors $g(v_\s{min})$ and $h(v_\s{min})$. By evaluating $g(v_\s{min})$ and $h(v_\s{min})$ for a large number of samples from the aforementioned MCMC scans, we obtained the posterior distributions of the astrophysical factors which allowed us to compute their median values as well as the corresponding credibility intervals. In comparison with the SHM, the axisymmetric models lead to astrophysical factors that are flatter at small $v_\s{min}$, have a pronounced knee at intermediate values of $v_\s{min}$ and fall to zero more rapidly as $v_\s{min}$ approaches the escape velocity. The differences between various axisymmetric models are typically small and there is a substantial overlap of their 68\% h.p.d. bands while the deviations in comparison with the SHM can be appreciably larger. Consequently, sizable differences can arise in the expected nuclear recoil rate, especially at the lower end of the DM mass range that can be probed in a given detector. There the cross-section limits inferred from the SHM can be overestimated by more than an order of magnitude. Smaller, but still notable, differences also arise at DM masses around the peak sensitivity of the detector, where the SHM yields up to $\sim 30\%$ weaker limits than the ones that can be set using the axisymmetric modeling. For even higher DM masses, the discrepancies between the considered models become less significant.
	
	Further important differences between the SHM and axisymmetric equilibrium models can arise in the attempts of reconstructing the DM properties and identifying the scattering operator upon successful detection of DM signals. To address these effects, we generated a large sample of stochastic realizations of observational datasets which were subsequently used to reconstruct the DM mass and coupling using MCMC sampling of the energy-binned likelihood function or identify the scattering operator responsible for the coupling based on the Bayesian evidence ratios using the same likelihood function. Focusing on the next-generation experiments that are currently being developed, namely the DARWIN and DarkSide-20k detectors, we found that for $m_\chi = 30 \, \s{GeV}$ and a cross-section slightly below the current most stringent limits the SHM performs reasonably well. It leads to a modest (few per cent) bias and slightly larger uncertainties in the reconstructed DM mass and coupling parameter for either SI or anapole scattering operators and the axisymmetric models perform better only if the correct rotational properties of the DM halo are assumed. As has been observed in several previous works, we confirm that for heavy DM the particle physics parameters become strongly degenerate regardless of the assumed astrophysical factors and can be disentangled only by using the results of multiple detectors with different target materials. However, upon combining the mock data of the DARWIN and DarkSide-20k detectors for $m_\chi = 100 \, \s{GeV}$ DM, we found that the SHM leads to more than $50\%$ larger uncertainties and $\mathcal{O}(10\%)$ bias in the central values of parameters when compared with the axisymmetric models. On the other hand, the differences among considered equilibrium models are much smaller, which demonstrates the benefit of using kinematically constrained axisymmetric PSDFs for computing the astrophysical factors. We found similar results for the prospects of identifying the scattering operator responsible for the interactions. While the differences are minor at $m_\chi \sim 30 \, \s{GeV}$, for $m_\chi = 100 \, \s{GeV}$ DM the analysis based on the SHM is more susceptible to Poissonian fluctuations in the energy-binned event rates in comparison to the trials based on axisymmetric models. While in the case of DARWIN measurements alone, it is difficult to identify the correct interaction type regardless of the assumed astrophysical factors, it becomes feasible when combining the DARWIN and DarkSide-20k detectors. In the latter case, the SHM yields a probability measure for the true operator of $\kappa > 0.79$ while the axisymmetric models lead to $\kappa > 0.93$ for 84\% of the stochastic realizations of the observational datasets. This demonstrates that the analyses based on axisymmetric models are more robust against statistical fluctuations in the event counts, which leads to more reliable identification of the interaction type.
	
	Further differences can also arise in other observables which are essential for rejecting various backgrounds and will become increasingly important as the sensitivity of detectors improves. First, we addressed the sizable variations in the expected amplitude of the annual modulation of the nuclear recoil rate. We found that the SHM can under-predict the modulation amplitude by up to 50\% in a significant portion of the accessible DM mass range, which is primarily due to its implicit assumption of the isotropic DM velocity distribution. Conversely, the axisymmetric models predict a significantly lower velocity dispersion along the azimuthal direction in comparison with the meridional plane, which leads to larger yearly variations in the expected recoil rate. Due to the same reason also subtle differences in the position and shape of the yearly minimum and maximum can occur, however, all of these effects become suppressed for sufficiently light or heavy DM.
	
	Finally, we also demonstrated the impact of axisymmetric models on the expected directional distribution of nuclear recoils. Our results suggest that the relative recoil rates can differ by more than 10\% in the direction with the highest number of expected events, while even more significant deviations can occur in dimmer regions. Furthermore, at large $v_\s{min}$ (corresponding to small DM mass or large recoil energies) the size of the sky patch from which the events are expected to originate can vary appreciably. While currently such information still has no significant influence on the interpretation of the direct detection experiments, it is expected to become crucial in the future when directional experiments reach sensitivities at which neutrinos from various astrophysical sources will become the dominant background.
	
	\paragraph*{Indirect detection:}
	
	After examining implications of the considered halo models for the direct detection experiments, we turned our attention to indirect searches of possible DM annihilation signals originating from the GC. In this context, the central DM distribution is particularly important, which leads to significant uncertainties due to the lack of reliable kinematic measurements in that region. The NFW and Burkert halos obtained from the MCMC sampling of the galactic mass models lead to largely inconsistent prediction for the astrophysical factors within the inner $10^\circ$ around the GC. While our results for the Burkert case can be taken as a lower bound on the $J$-factor, an even steeper central cusp than the $\rho \propto r^{-1}$ of the NFW density profile could be present in our galaxy, leading to even larger values of $J$. On the other hand, we identify a circular band centered at roughly $20^\circ$ around the GC that is characterized by an astrophysical factor which is significantly less dependent on assumptions regarding the central slope of the DM density profile. In indirect searches further notable differences between the considered PSDF models can arise if the DM annihilation cross-section exhibits non-trivial velocity dependence, which occurs in case of, e.g., p-wave or Sommerfeld enhanced annihilations. Guided by these considerations, we evaluated the first two positive and negative moments of the relative velocity distribution for the DM particles as determined by our benchmark PSDF models. As could be expected, the differences between Maxwellian approximation and equilibrium models are small at large galactocentric distances, however, they become substantial in the proximity of the baryonic disc. There the axisymmetric models tend to predict $\mathcal{O}(10\%)$ higher (lower) values of positive (negative) relative velocity moments along the disc plane, resulting in a slightly enhanced (suppressed) $J$-factor in comparison with the corresponding model based on the Maxwell-Boltzmann velocity distribution. Within the inner kpc, the difference can become even more significant, however, there the most important uncertainties are sourced by the poorly constrained central slope of the DM density profile. Since cuspy and cored models lead to diverging predictions for both, equilibrium and Maxwellian, velocity distributions this results in large systematic errors in the moments of the relative velocity. Furthermore, it should be noted that some caution is needed in extrapolating our results to the very central region of the galaxy -- besides not being constrained by the observations, the galactic bulge exhibits a triaxial shape which cannot be captured within an axisymmetric framework. Despite these shortcomings, our results still provide a valuable re-evaluation of the astrophysical factors associated with the GC.
	
	In conclusion, our work provides novel results regarding the equilibrium phase-space structure of the Milky Way's DM halo. By adopting a data-driven approach within the Bayesian framework, we were able to consistently propagate the observational uncertainties throughout the analysis and obtain robust credibility intervals on various astrophysical factors that are crucial for the interpretation of DM searches. It has been demonstrated that the axisymmetric models can lead to significant differences in comparison with the predictions of commonly used Maxwellian halo models, especially in the context of direct detection. However, it is important to note that our models are valid only for the component of the galactic DM that has reached dynamical equilibrium, while important corrections can arise from structures that have not yet fully phase-mixed. The latter are generically expected to exist in galaxies, e.g., in the form of tidal debris or gravitationally bound remnants from past mergers, but should represent a subdominant component of the DM halo. Therefore, they can be accounted for as additional contributions to the DM's phase-space distribution function, added on top of the accordingly re-scaled equilibrium models, as has been done in a number of recent studies. A more challenging drawback of our approach is the fact that it does not allow us to account for non-equilibrium dynamics, such as the response of the galactic DM halo to external gravitational perturbations caused by the nearby Magellanic Clouds. However, studying such effects is notoriously difficult and can be, to the best of our knowledge, addressed only through dedicated numerical simulations. Further improvements of our models could also be achieved by dropping the assumption of axial symmetry and adopting fully general action-angle modeling. In this context, our results could serve as a starting point for constructing more elaborate models which are, on the other hand, often very difficult to constrain through the existing observations and require much greater computational resources. There are also several possible improvements regarding the derived astrophysical factors. In the context of direct detection, one should ideally marginalize the predictions of axisymmetric models over the observationally unconstrained rotational properties of the Milky Way's halo which currently represent an irreducible source of systematic uncertainty. Some insight concerning the halo rotation could be perhaps obtained from novel findings regarding the merger history of our galaxy and comparing it with cosmological simulations. However, since this represents a frontier in our understanding of the galactic evolution, we postpone such a detailed analysis to future work. Concerning indirect detection, important improvements could be achieved by accurately assessing the systematic uncertainties related to the assumed parametric forms of the DM density profile. Furthermore, it would be highly desirable to properly constrain the central DM density slope, however, this would require significantly more accurate observational constraints on the kinematics within the inner few kpc of the Milky Way. Finally, we plan to assess the accuracy of axisymmetric inversion method by comparing its predictions with the actual phase-space distribution within simulated galaxies. This upcoming work will also allow us to establish the advantages and shortcomings of the axisymmetric approach in comparison with the more general action-angle models.

	\section*{Acknowledgments}
	
	Foremost, I would like to thank Piero Ullio for motivating the preparation of this work and providing his guidance. Further credit goes to Benoit Famaey and Fabrizio Nesti for their help and suggestions. I would also like to thank Julien Lavalle and Vivian Poulin for stimulating debates. Furthermore, I am very grateful to the GAMBIT Collaboration for providing the DDCalc software and, in particular, to Felix Kahlhoefer for his technical support. Finally, I would also like to thank the referees for providing very useful comments and corrections during the peer-review process. The author acknowledges partial support from the European Union's Horizon 2020 research and innovation program under the Marie Sk\l odowska-Curie grant agreements No 690575 and 674896, as well as the ANR project ANR-18-CE31-0006.
	
	\bibliographystyle{unsrt_man}
	\bibliography{MW_local_DM}
	
\end{document}